\documentclass[iop]{emulateapj}
\usepackage{graphicx}
\usepackage{epstopdf}
\usepackage{natbib}
\usepackage{amsmath}

\providecommand{\e}[1]{\ensuremath{\times 10^{#1}}}
\usepackage{url}

\shortauthors{Hawley et al.}
\shorttitle{Kepler Flares I.}

\begin{document}
\title{Kepler Flares I. Active and Inactive M dwarfs}

\author{Suzanne L. Hawley\altaffilmark{1,2}, 
James R. A. Davenport\altaffilmark{2},
Adam F. Kowalski\altaffilmark{2,3}, 
John P. Wisniewski\altaffilmark{2,4},\\
Leslie Hebb\altaffilmark{5},
Russell Deitrick\altaffilmark{2},
Eric J. Hilton\altaffilmark{2,6}} 

\altaffiltext{1}{Corresponding author: slhawley@uw.edu}
\altaffiltext{2}{Department of Astronomy, University of Washington, Box 351580, Seattle, WA 98195}
\altaffiltext{3}{NASA Goddard Space Flight Center, Code 671, Greenbelt, MD 20771}
\altaffiltext{4}{HL Dodge Department of Physics \& Astronomy, University of Oklahoma, 440 W. Brooks Street, Norman, OK 73019}
\altaffiltext{5}{Department of Physics, Hobart and William Smith Colleges, 300 Pulteney Street, Geneva, NY 14456}
\altaffiltext{6}{Universe Sandbox, 911 E. Pike Street \#333, Seattle, WA 98122}

\begin{abstract}
We analyzed Kepler short-cadence M dwarf observations. Spectra from the ARC 3.5m telescope identify magnetically active (H$\alpha$ in emission) stars.  The active stars are of mid-M spectral type, have numerous flares, and well-defined rotational modulation due to starspots.  The inactive stars are of early-M type, exhibit less starspot signature, and have fewer flares.  A Kepler to U-band energy scaling allows comparison of the Kepler flare frequency distributions with previous ground-based data.  M dwarfs span a large range of flare frequency and energy, blurring the distinction between active and inactive stars designated solely by the presence of H$\alpha$.  We analyzed classical and complex (multiple peak) flares on GJ 1243, finding strong correlations between flare energy, amplitude, duration and decay time, with only a weak dependence on rise time.  Complex flares last longer and have higher energy at the same amplitude, and higher energy flares are more likely to be complex.  A power law fits the energy distribution for flares with log $E_{K_p} >$ 31 ergs, but the predicted number of low energy flares far exceeds the number observed, at energies where flares are still easily detectable, indicating that the power law distribution may flatten at low energy. There is no correlation of flare occurrence or energy with starspot phase; the flare waiting time distribution is consistent with flares occurring randomly in time; and the energies of consecutive flares are uncorrelated.  These observations support a scenario where many independent active regions on the stellar surface are contributing to the observed flare rate.  
\end{abstract}

\section{Introduction}

The Kepler satellite \citep{borucki2010} has ushered in a new era of 
stellar photometric 
investigation, enabling light curve analysis with unprecedented precision:
approaching 10 ppm in bright targets (V=9-10) and 100 ppm
even in faint targets (V=13-14).  Light curves of this precision,
over
periods of months to years with minimal interruption, provide a unique view of
stellar variability in optical light.  Here, we are interested in
the variability that results from flares on low mass stars.

The advantages of Kepler over ground-based observing of stellar flares
are many.  Since flares occur at unpredictable intervals, a long time baseline
uninterrupted by daylight, weather or the vagaries of telescope scheduling 
greatly facilitates the determination of
the flare frequency distribution (FFD, the cumulative number of flares above a given energy that occur per unit time).  FFDs on low mass stars have been extensively examined from 
the ground, beginning with the pioneering work of \citet{gershberg1972} and 
the seminal paper by \citet[][hereafter LME]{lme1976} which presented results for 8 stars 
using more than 400 hours (spread over two years) of
ground-based monitoring observations by \citet{moffett1974}.  These early studies showed that 
flares occur more frequently but with lower energy on mid-M dwarfs 
(M3-5)  compared to earlier type, M0-M2 stars.  However, there are significant 
selection effects at both extremes of the distribution due to 
a) the relatively higher quiescent luminosity of the earlier type stars, 
which makes it more difficult to detect low energy flares; and 
b) the limited duration of the monitoring observations which makes it 
less likely to observe the relatively rare, high energy flares.  Recently, 
\citet{hiltonthesis} compiled several hundred hours of ground-based monitoring 
of M dwarfs and extended the investigation of FFDs to the regimes of very 
low mass (VLM) dwarfs of type M6-M8 and of inactive, early type M dwarfs.  
His results indicate that the trend of higher frequency but lower energy 
flares extends to the latest M dwarfs, and that even ``inactive'' 
(no H$\alpha$ emission in the quiescent spectrum) early-type 
M stars still flare, although at much lower frequency than active stars 
of the same spectral type.  \citet{audard2000} presented FFDs for several M dwarfs using X-ray data from the EUVE satellite and obtained similar results to the optical studies.  HST investigations in the NUV \citep{robinson1995} and FUV \citep{robinson2001} revealed smooth extension of the FFDs to the microflare regime for both early and late M dwarfs.
In the first reported long-duration optical 
monitoring observations from space, \citet{huntwalker2012} used data
from the MOST satellite on the active M3 star AD Leo to determine that
the FFD from a week of nearly continuous observation matched those previously
found from the ground.

With a power-law probability distribution for flare energies
\begin{equation}
N(E) dE \propto E^{-\alpha} dE,
\end{equation}
a linear fit to the cumulative FFD gives a slope $-\alpha + 1$.  There is considerable interest in the solar literature concerning the value of $\alpha$, since a flare energy distribution with $\alpha > 2$ extending to low flare energies would potentially provide enough energy from flares to heat the solar corona \citep{parker1988,hudson1991,hannah2011,schrijver2012}.  The M dwarf studies described above typically find flatter power law distributions (smaller $\alpha$) than for the Sun, at least for ground-based data.   \citet{gudel2004} reviewed X-ray results and noted several studies that obtain $\alpha >2$.   \citet{ramsay2013} recently presented FFDs from Kepler data for two M dwarfs (including the active M4 star GJ 1243 which is contained in our sample), and found steeper power-law slopes, also implying $\alpha > 2$.  By contrast, the Kepler study of an early L dwarf examined by \citet{gizis2013} showed a relatively shallow power-law slope in its FFD.  

The assurance that every flare has been observed over a long time period of consecutive
observation (approximately one month with Kepler) also
provides excellent statistics for the study of flare timing, for example the waiting time between 
flares. 
The waiting time distribution can be used to address whether flares occur independently, or are triggered by other flares (sympathetic flaring), and also whether the phenomenon of precursor flares (small flares immediately followed by large flares) is real.  LME found that the flares in their sample were consistent with random occurrence, but other studies have presented evidence for sympathetic flaring on low mass stars \citep{pazzani1981,doyle1990,panagi1995}, RS CVn stars \citep{osten1999}, and on the Sun \citep{pearce1990}.  \citet{wheatland2010} investigated flares within a single solar active region, and found that they followed a piecewise Poisson distribution, indicating random occurrence, but with different rates as the magnetic topology of the region evolved.

In addition, the precision of the Kepler light curves provides an unprecedented view of periodic
variations attributed to starspots rotating across the visible hemisphere.
Starspots on M dwarfs have previously been difficult to measure
due to the very small amplitudes ($< 0.01$ mag) that are typically 
observed \citep{irwin2011}.  However, variations of this amplitude are easily measured with 
Kepler.  It is well known on the Sun that flares are associated with
active regions (sunspots).  The Kepler data now make it possible to look for a correlation between
starspot phase and flare occurrence in low mass stars.  

Finally, since Kepler monitors many stars 
simultaneously, it is possible to compare these various measures of flare
energy, frequency and timing in stars of different mass, temperature and
evolutionary state as well as in otherwise similar stars
that exhibit very different levels of magnetic activity.  A series of papers
reporting spectropolarimetric observations of M dwarfs \citep{morin2008b,donati2008,morin2010}
describes the change in the large-scale magnetic field structure between early M dwarfs that retain a radiative core, presumably have a solar-type dynamo and exhibit weak, mostly toroidal field; and mid M dwarfs, which are
fully convective and have a turbulent dynamo, and exhibit stronger, more stable, poloidal fields.
As these authors point out, the large scale fields they infer from their spectropolarimetry have relatively low field strength and most of the magnetic energy in the active stars, with measured field strengths of several kG \citep{cmj1996}, must be contained in smaller, spatially confined active regions that are unresolved with global polarization measurements.  \citet{reiners2009} reported Stokes V and I measurements of early and mid M dwarfs and find that the large scale fields contribute at most 15\% of the total magnetic flux.
Investigation of flare activity in both active and inactive stars of early and mid M dwarfs may provide additional insight into the relative importance of the large and small scale fields, especially at spectral types near M4, where stars transition to a fully convective interior and turbulent dynamo
(see discussion in \citet{stassun2011}).

We have carried out a long term program of flare monitoring with Kepler in the short (one-minute) cadence mode, with the ultimate goal of obtaining a deeper understanding of the flare mechanism, how and when flares occur and how flare properties are correlated with the underlying stars.
This is the first in a series of papers describing our results.  In this
paper (Paper 1), we focus on the lowest mass stars in our Kepler program, active and inactive M dwarfs
of spectral type M1-M5.   \citet[][hereafter Paper 2]{davenport2014b} examines flare light curve morphology using a large sample of flares on the active M dwarf GJ 1243.  Lurie et al (2014 in prep., hereafter Paper 3), investigates the active binary system GJ 1245AB using the Kepler pixel files to separate the two stars and analyze the individual light curves.  Additional papers are also planned.

Section 2 of this paper describes the M dwarf sample and the process we use to identify flares.  
Section 3 presents the results for flare frequencies and energies, and
the comparison between the active and inactive stars.  Sections 4 and 5 concentrate on 
the active single star GJ 1243, and describe the relationships between basic properties such
as flare amplitude, duration and energy, as well as our analysis of flare timing and correlation with starspot phase.  We offer some conclusions and discussion of future work in Section 6.

\section{Kepler M Dwarf Flare Sample}

\begin{figure*}[]
\centering
\includegraphics[width=7in]{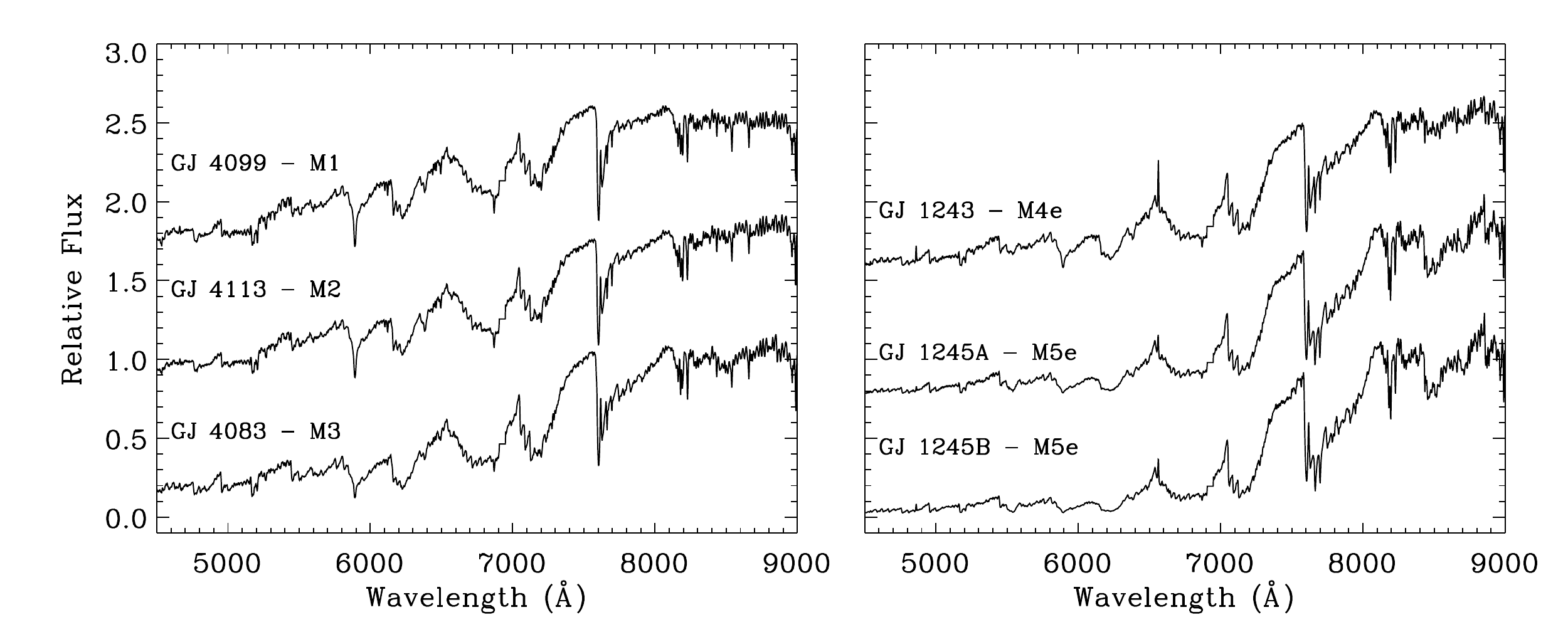}
\caption{Moderate resolution spectra obtained with the 
ARC 3.5m Telescope and DIS spectrograph for the M dwarfs
in the Kepler flare sample.}
\label{fig:spec}
\end{figure*}

\begin{figure*}[]
\centering
\includegraphics[width=7in]{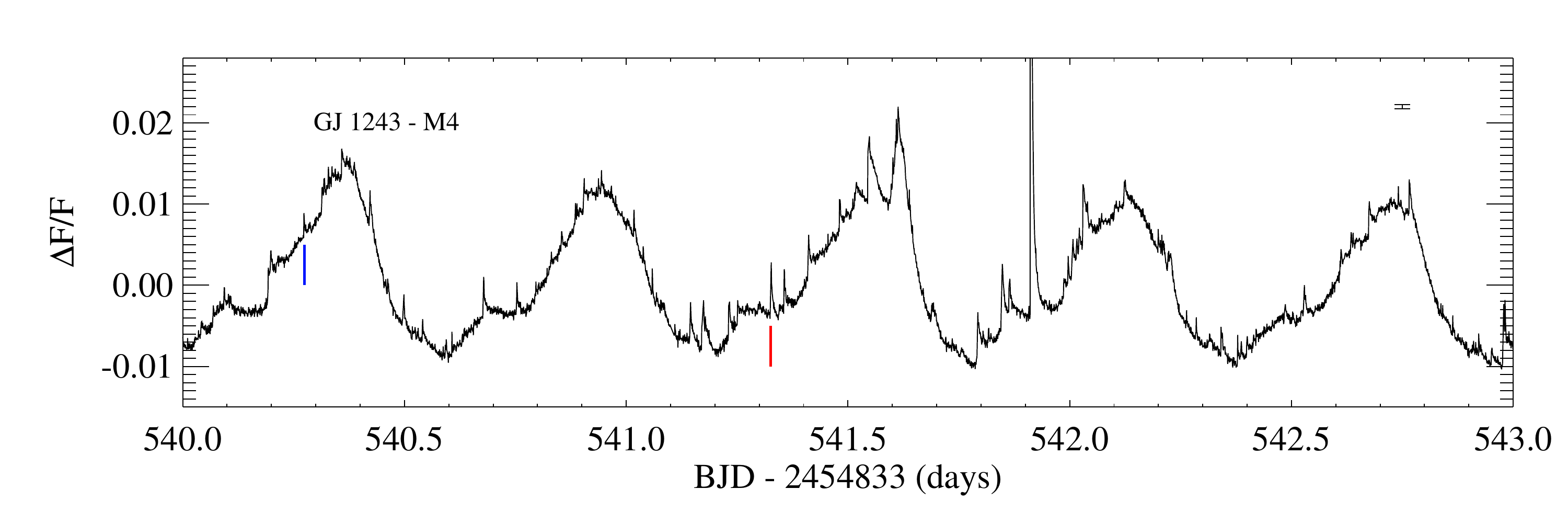}\\
\includegraphics[width=7in]{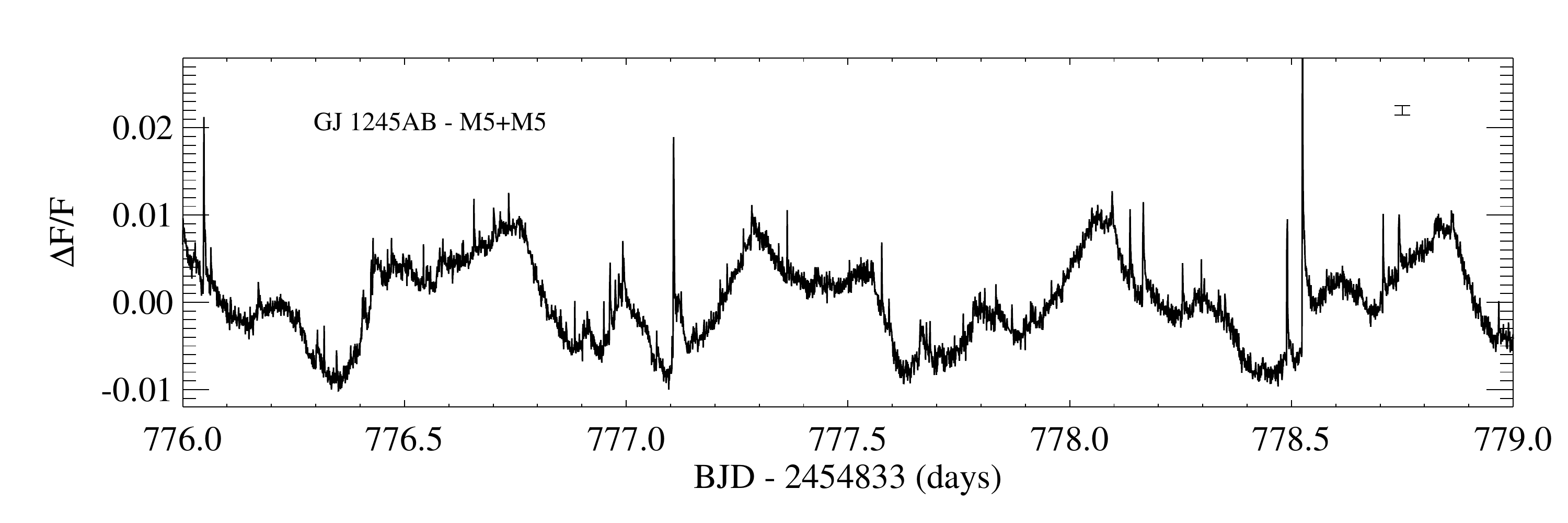}
\caption{Light curves for the active M dwarf GJ 1243 (top) and the active 
binary GJ 1245AB (bottom).  A flare with energy log $E_{Kp}\sim$30.5 is marked with a vertical blue bar, and a flare with energy log $E_{Kp}\sim$31 is marked with a vertical red bar; these are discussed in \S 4.1.
Although GJ 1245AB is brighter, it exhibits
more noise likely due to variations in the amount of total light from the
binary falling within the Kepler photometric aperture (see text).}
\label{fig:activelc}
\end{figure*}

\begin{figure*}[]
\centering
\includegraphics[width=7in]{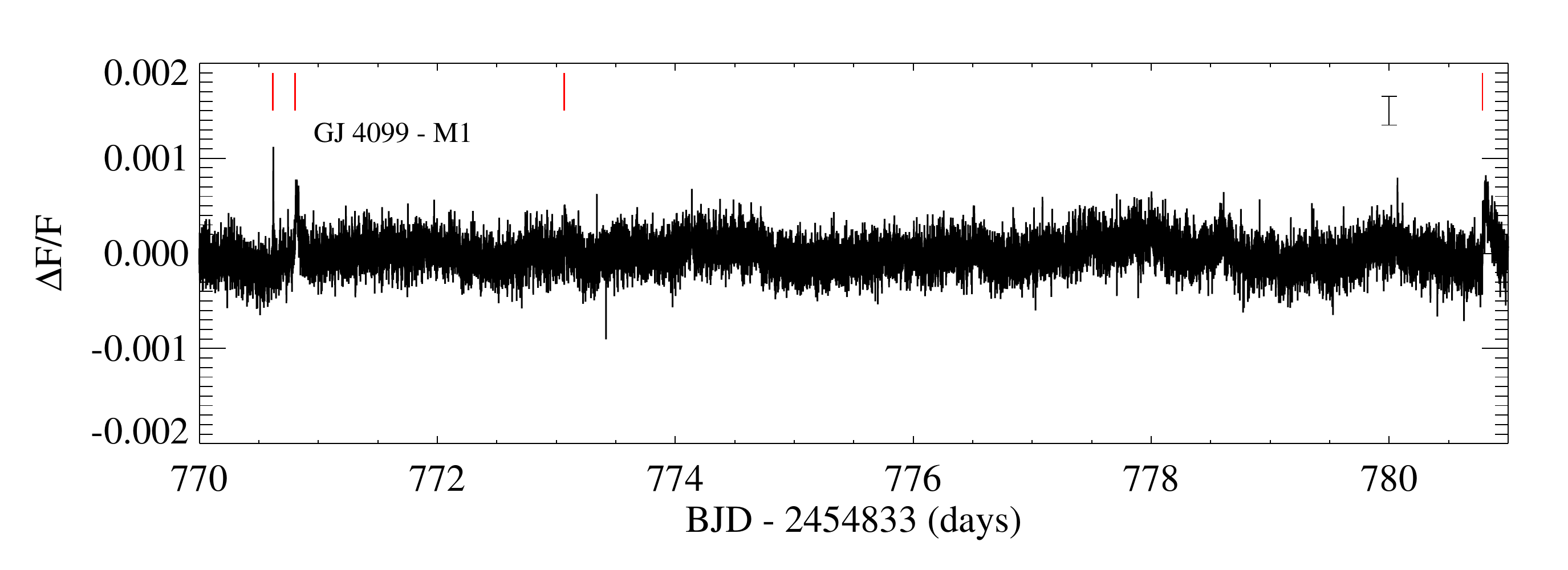}
\includegraphics[width=7in]{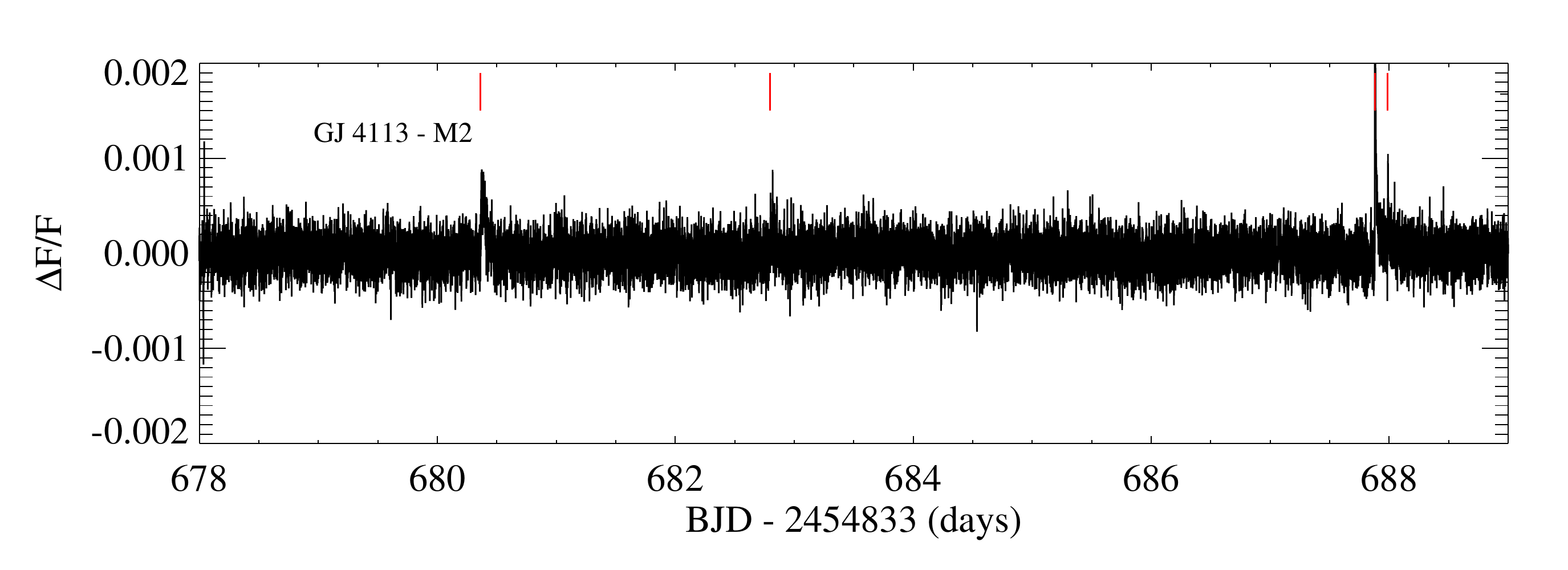}
\includegraphics[width=7in]{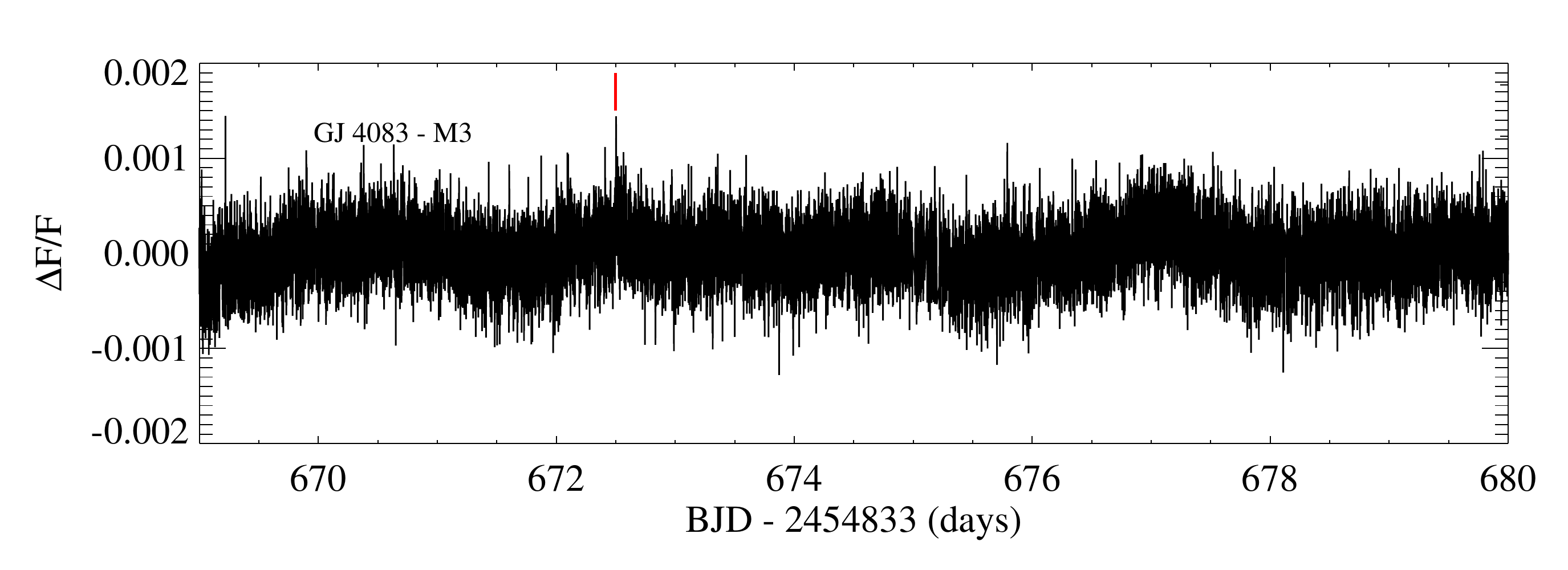}
\caption{Light curves for the inactive M dwarfs GJ 4099 (M1), GJ 4113 (M2) and GJ4083 (M3).  Note the increased noise for GJ 4083, which is 
more than a magnitude fainter than the other two inactive stars.  Flares that were identified in
these light curve segments are marked with vertical red lines.}
\label{fig:inactivelc}
\end{figure*}

We chose the most active and brightest late type M dwarfs 
in the Kepler field, GJ 1243 (dM4e)
and GJ 1245AB (dM5e+dM5e) and the latest type inactive 
(no H$\alpha$ emission) M dwarfs 
(GJ 4083, dM3; GJ 4099, dM1; and GJ 4113, dM2) that were bright 
enough to give good signal in one-minute cadence data.  To our knowledge,
all of the objects are single field stars, except for the 
GJ 1245AB system which is comprised of two nearly identical active 
mid-M dwarfs separated by 7'', and a third component (GJ 1245C), 
an M8 companion to GJ 1245A, separated by 0\farcs6 \citep{schroeder2000}. 
The entire system is measured together in the Kepler pipeline photometry.  
Since GJ 1245C is 
approximately 3 magnitudes fainter than the GJ 1245AB system, we do not 
expect that it contributes significantly to the combined light observed with
Kepler.  In particular, a 3 magnitude flare in the Kepler bandpass on an M8 dwarf
will occur less than once every 100 hours according to results from \citet{davenport2012}
and \citet{hiltonthesis}, and such flares would only appear as brief, short-lived, low-energy excursions in the GJ 1245AB light curve.  We therefore treat GJ1245AB as a binary system and analyze the light
curve in terms of the A and B components.  In Paper 3 of this series 
we analyze the GJ 1245 system in more detail and separate the components using recalibrated light curves from the Kepler pixel files.

Table 1 gives the basic properties of the sample targets, including our 
determination of rotation periods which we describe below.
We obtained new spectral types using moderate resolution spectra obtained with
the DIS spectrograph on the ARC 3.5m telescope at Apache Point Observatory. 
Figure \ref{fig:spec} shows the six spectra (GJ 1245 A and B were resolved so 
that we were able to take spectra of the individual components).  
The spectral typing followed the
procedure of \citet{covey2007} using the Hammer software described 
therein.  Distances for the inactive stars in Table 1 were determined using 
published $R-I$ colors transformed to SDSS $r-i$ \citep{davenport2006}
and then employing
the $M_r$ vs $r-i$ photometric parallax relation given in 
\citet{bochanski2010}, with the updated values given in the erratum to that paper \citep{bochanski2012}.
The active stars have measured parallaxes as reported in \citet{pmsu1}.

\subsection{Kepler Light Curves}
Our Kepler Cycle 2 GO program 20016
comprised two months of one-minute cadence data for each of the objects
in the sample.  Note that the one-minute 
cadence is important to obtain well-sampled flare light curves, since
many flares on M dwarfs have a duration shorter than 30 
minutes \citep[e.g.][]{moffett1974} and nearly all are shorter than the 
90 minutes (3 observations of 30 minutes each) required in e.g. the 
\citet{walkowicz2011} discussion of stellar flares in Kepler long 
cadence data.  

We used the most recent publicly available version of the Kepler photometry, which gives
the Pre-search Data Conditioning Simple Aperture Photometry (PDCSAP) 
flux \citep{smith2012}. The PDCSAP fluxes are the result of a systematic 
reprocessing of the entire Kepler database using a Bayesian approach to 
remove systematics from the short and long term light curves. However, 
some month to month corrections still needed to be manually applied. 
We used a linear least-squares fit to each month of data to remove
low-order flux variations, and normalized the median flux levels between 
months. 

Figure \ref{fig:activelc} illustrates representative 3-day segments from the 
light curves for the active stars, with the fractional flux $\Delta F/F$ given by:
\begin{equation}
\Delta F/F = (F_i - F_o)/F_o 
\end{equation}
where $F_i$ is the measured flux sampled at each time $i$, and $F_o$ is the normalized median flux as described above.

The light curves display a simple, well-defined, 
periodic flux variation in GJ 1243 and a more complicated
flux variation for the binary GJ 1245AB, due to Kepler measuring the 
combined light from both stars.  The periodic flux variations are 
ascribed to starspots moving in and out of view on the visible hemisphere as the star rotates.  These starspot variations are expected on magnetically active stars, and have been observed commonly on active G and K stars even from the ground, with flux variations up to 10\%.  M dwarfs exhibit much smaller flux variations due to starspots; even the very active stars only show variations of  $\sim$1\%.  If much of the star is covered with strong magnetic fields, as discussed in \citet{cmj1996},
then numerous small (or a few large) spots may always be present on the surface, and thus only small variations as the star rotates would be expected.
  
Many flares are evident even in these 
short data segments.  For reference, flares with energies in the Kepler bandpass (see \S 2.3 for discussion of energy measurements) of log $E_{Kp}$ = 30.5 and 31 are indicated with vertical blue and red bars respectively.  These particular flare energies will be discussed further in \S 4.1 below.

The increased noise in the GJ 1245AB light curve 
compared to GJ 1243 is unusual since GJ 1245AB is a brighter target.  
However, inspection of the pixel files for GJ 1245AB reveal that it 
is partially resolved, with much of the light from the A component in particular
falling outside the aperture chosen for the pipeline photometry.  Additional noise results from
varying amounts of light from the two stars falling within the Kepler 
photometric aperture due to spacecraft jitter and other guiding issues.
These issues also affect the flux normalization, flare identification and flare energies
assigned to GJ 1245AB, adding uncertainty to the results obtained from the combined
system.

Figure \ref{fig:inactivelc} shows 11-day segments of the inactive star 
light curves, indicating far less flux variation due to starspots (note the flux scale is
a factor of ten reduced compared to Figure 2) and a few small flare events, marked with red vertical bars.  It is notable
that even in these inactive (no H$\alpha$ emission) stars, we still see evidence for some magnetic activity from starspots and flares.

\begin{figure}[]
\centering
\includegraphics[width=3.4in]{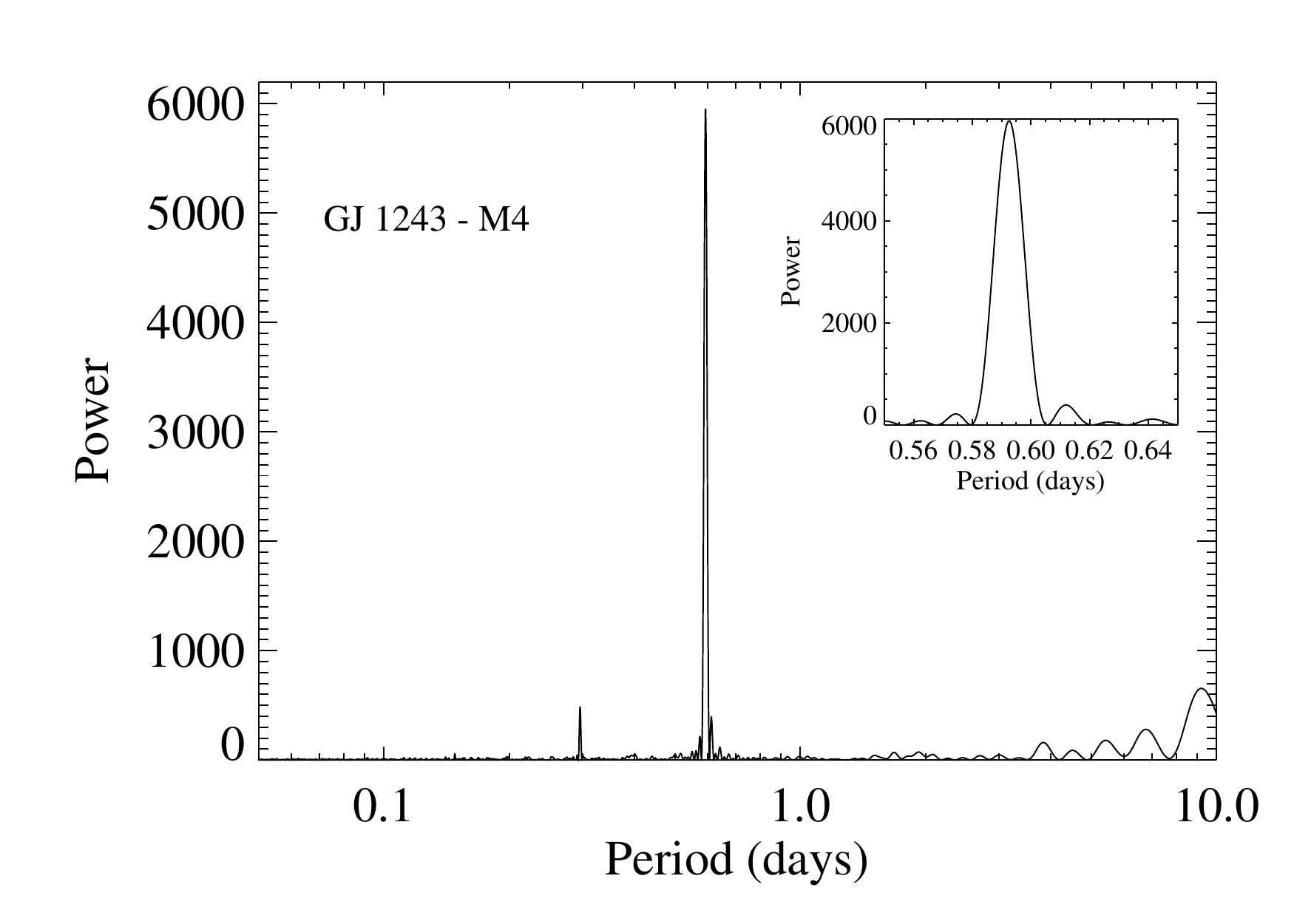}
\includegraphics[width=3.4in]{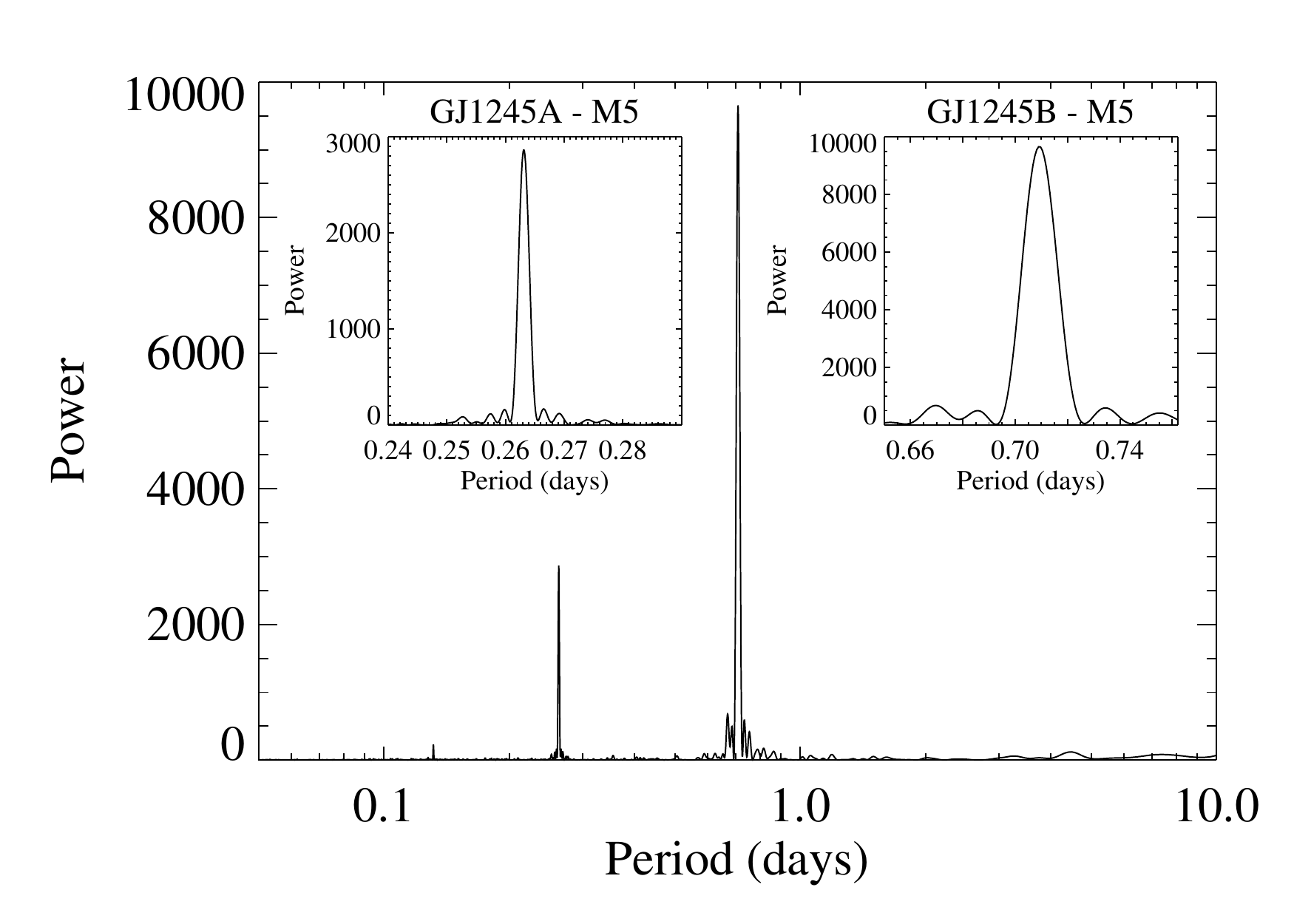}
\caption{Lomb-Scargle periodograms obtained from the Kepler light 
curves for the active star GJ 1243 (top) and the active binary GJ 1245AB (bottom).  The insets
show expanded views of the strongest peaks.}
\label{fig:per}
\end{figure}

\begin{figure}[]
\centering
\includegraphics[width=3.4in]{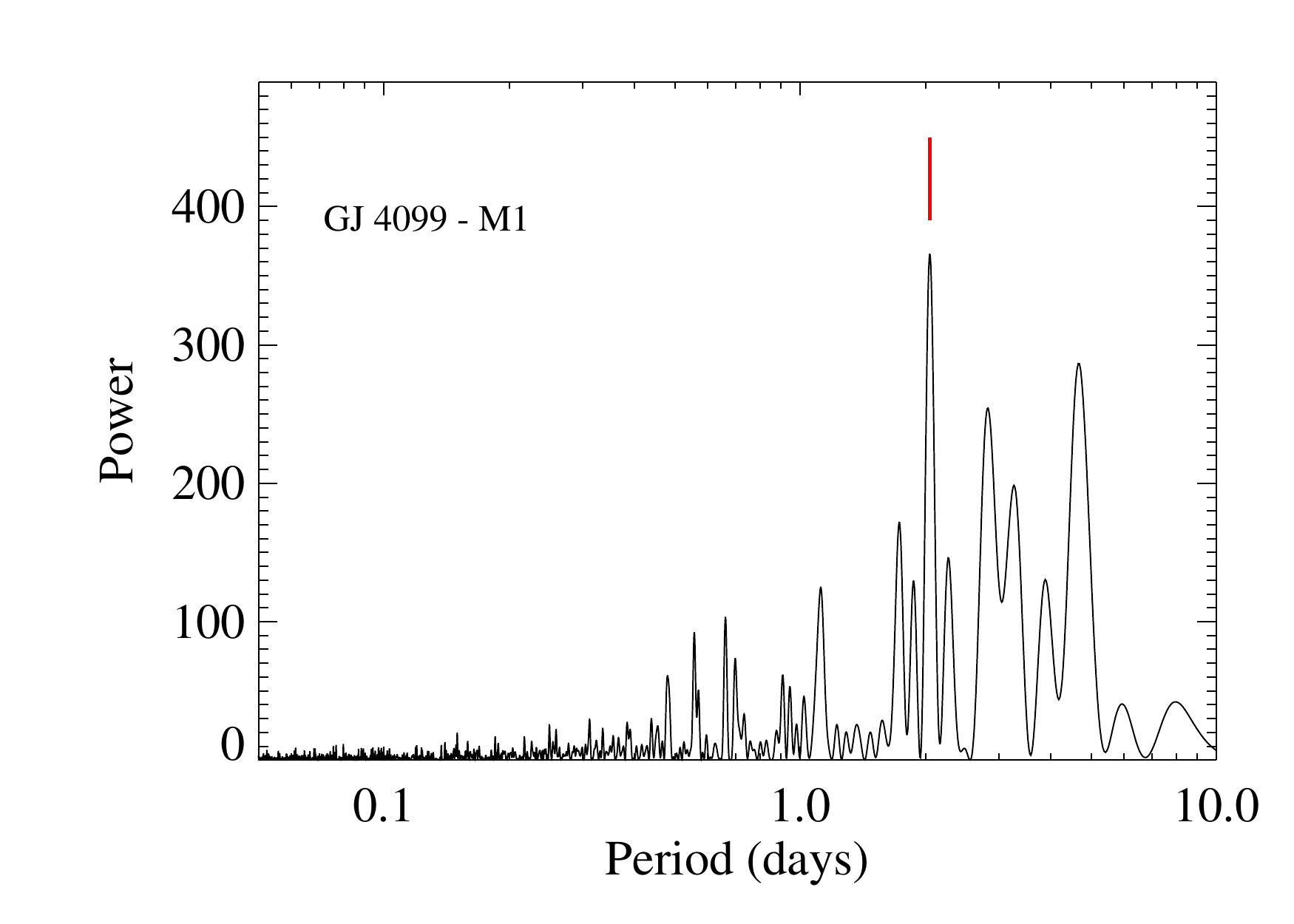}
\includegraphics[width=3.4in]{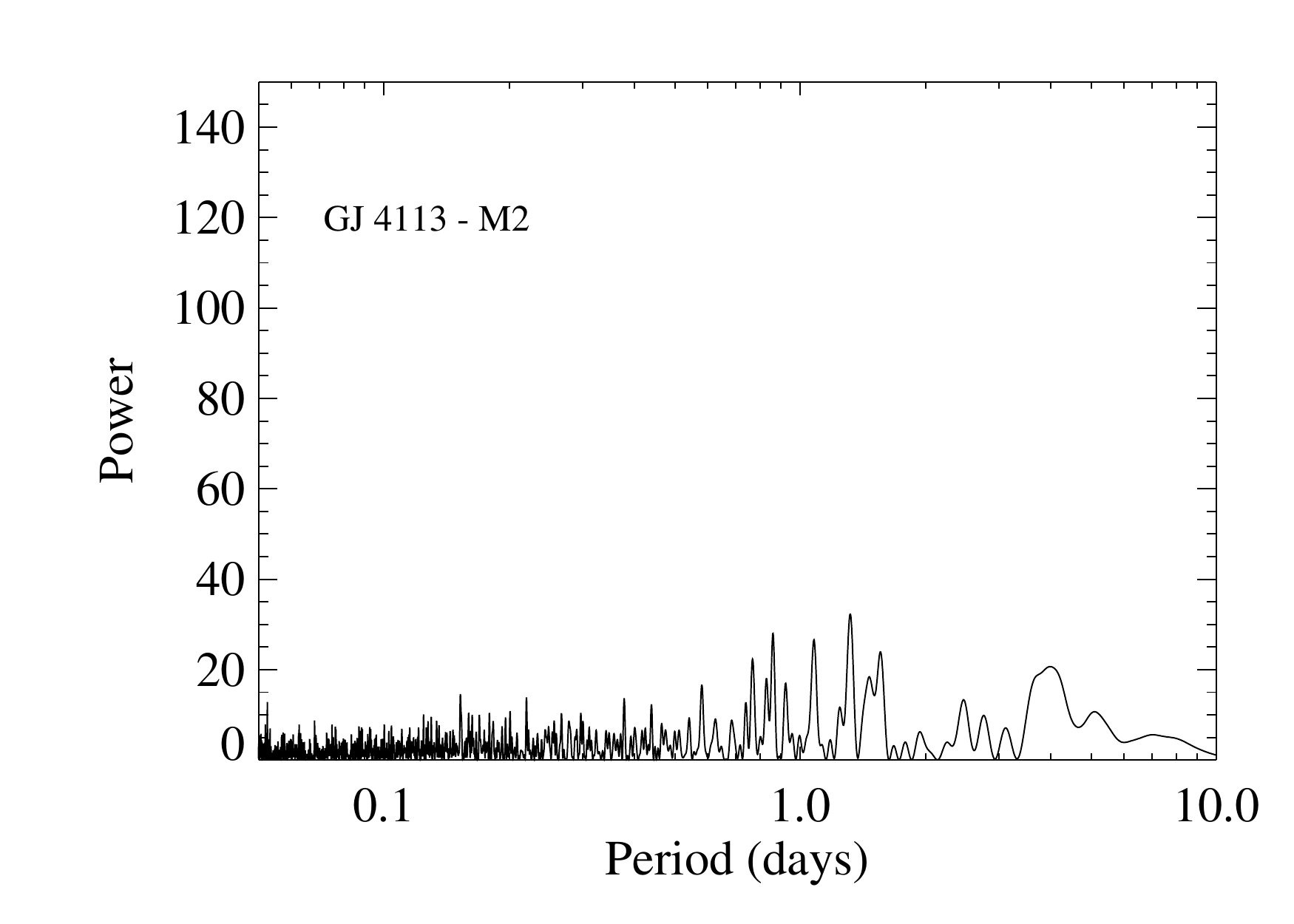}
\includegraphics[width=3.4in]{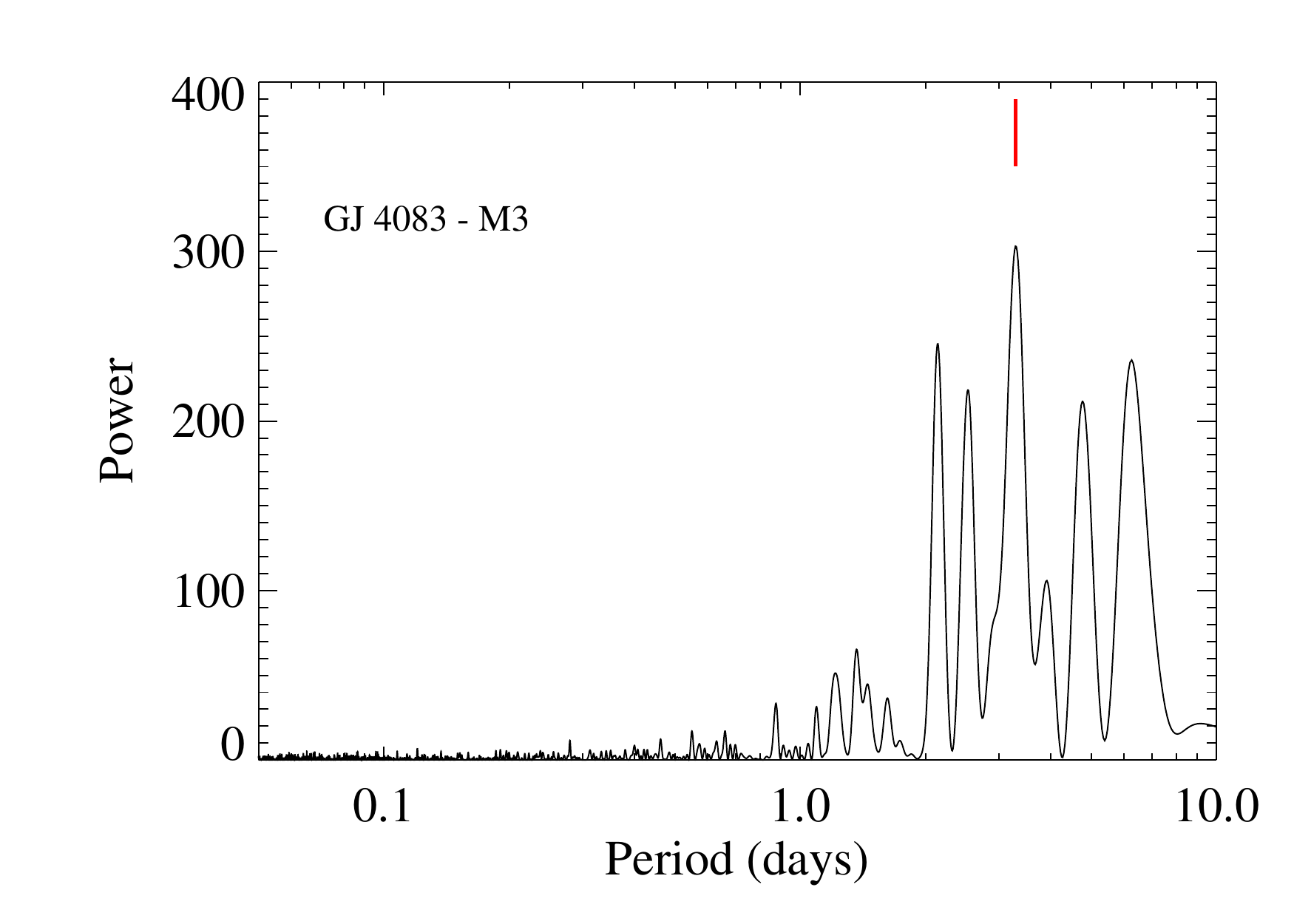}
\caption{Lomb-Scargle periodograms obtained from the Kepler light curves for the inactive 
stars GJ 4099 (M1, top), GJ 4113 (M2, middle) and GJ 4083 (M3, bottom).  The peaks representing the  periods reported in Table 1 for GJ 4099 and GJ 4083 are indicated.  We were unable to 
determine a period for GJ 4113. }
\label{fig:inactiveper}
\end{figure}

Figures \ref{fig:per} and \ref{fig:inactiveper} give the corresponding Lomb-Scargle periodograms for the 2-month Kepler dataset for each target.
In Figure \ref{fig:per},
GJ 1243 (P=0.59 days) and GJ 1245AB (P = 0.26 days for component A and P = 0.71 days for component B) have very strong
peaks in their periodograms indicating clearly periodic variations which
we attribute to rotational modulation from stable starspots on the
surfaces of these stars.  Our periods for these stars are reported in Table 1, with uncertainties found from comparing values obtained when analyzing separate months of data (the formal uncertainties on the individual fits are much smaller).  Our periods agree with previous determinations for GJ 1243 \citep{irwin2011,mcquillan2013} and GJ 1245B \citep{morin2010}.
To our knowledge this is the first measurement of the period of GJ 1245A.   Note that the combined Kepler pipeline light curve for GJ1245AB does not 
allow us to independently identify which period belongs to which star, but our additional pixel-level investigation in Paper 3 verifies that the periods are ascribed to the correct components, and provides the uncertainties quoted on those periods.

\begin{deluxetable*}{lccccccc}
\tablecolumns{8}
\tablecaption{Properties of the Kepler M dwarf flare sample}
\tablehead{
	\colhead{Star}&
	\colhead{KIC}&
	\colhead{Sp Type} &
	\colhead{$m_{Kp}$} &
	\colhead{Active} &
	\colhead{EW H$\alpha$ (\AA)} &
	\colhead{Dist (pc)} &
	\colhead{Period$^\dagger$ (d)}
	}
\startdata
GJ 4099 & 004142913 & M1 & 10.9 & n & \ldots& 16.2$^\star$ & 2.0:\\
GJ 4113 & 004470937 & M2 & 10.9 & n & \ldots& 14.2$^\star$ & \ldots\\
GJ 4083 & 010647081 & M3 & 12.3 & n & \ldots& 18.2$^\star$ & 3.3:\\
GJ 1243  & 009726699 & M4e & 12.7 & y & 4.0 & 12.05 & {0.5927 $\pm$ 0.0002}\\
GJ 1245A & 008451881 & M5e & 11.7$^+$ & y & 2.5 & 4.55 & {0.2632 $\pm$ 0.0001}\\
GJ 1245B & 008451881 & M5e & 11.7$^+$ & y & 3.8 & 4.55 & {0.709 $\pm$ 0.001}
\enddata
\tablenotetext{$\star$}{Distances determined via photometric parallax}
\tablenotetext{+}{combined magnitude for both components}
\tablenotetext{$\dagger$}{periods determined in this paper, see \S 2.1}
\label{datatable}
\end{deluxetable*}

The inactive stars GJ 4083 
and GJ 4099 (see Figure \ref{fig:inactiveper}) have several peaks of about the same weak 
power (an order of magnitude weaker than in the active stars) in their 
periodograms. These may be due to small transient starspots appearing 
on the stellar surface, possibly at different latitudes which may indicate
differential rotation.  We report the period of the strongest peak in Table 1 for GJ 4083 and
GJ 4099 (as indicated by red vertical lines in Figure \ref{fig:inactiveper}), 
but the periods are very uncertain, and should only be 
regarded as suggestive.  
The final inactive star, GJ 4113, does not show any significant peak in
its periodogram, and we are unable to determine its rotation period.  We also note that our analysis of two months of short cadence data is sensitive only to periods less than 10 days.  Previous examination of long cadence Kepler data revealed marginal evidence for longer periods in
the inactive stars in our sample \citep{mcquillan2013}.

\subsection {Flare Identification}
In order to identify flares, 
we used a modified version of the flare-finding approach described
in \citet{hiltonthesis} and \citet{huntwalker2012}. A full description of our flare finding methodology is given in Paper 2, and we briefly summarize the algorithm here. 
Each month of Kepler data was treated independently. 
The light curve was first de-trended to remove starspot modulations using a variable-span 
smoothing algorithm, similar to the ``Supersmoother'' method \citep{supersmoother}.
The local mean and standard deviation ($\sigma$) of the flux were then computed 
at each time sampled,
and candidate flares were flagged as excursions of two or more consecutive 
points more than 2.5$\sigma$ above the local mean flux.   Each month of data was independently inspected by at least two people, and both had to agree that a candidate flare was real for it to be included in the sample.    A flare type (classical or complex) is assigned based on whether there is only one peak in the light curve (classical) or more than one peak (complex), and these types were also manually verified.  Finally, the assigned start and end times for each flare were examined, and in some cases modified, during the inspection process.

\begin{figure}[]
\centering
\includegraphics[width=3.4in]{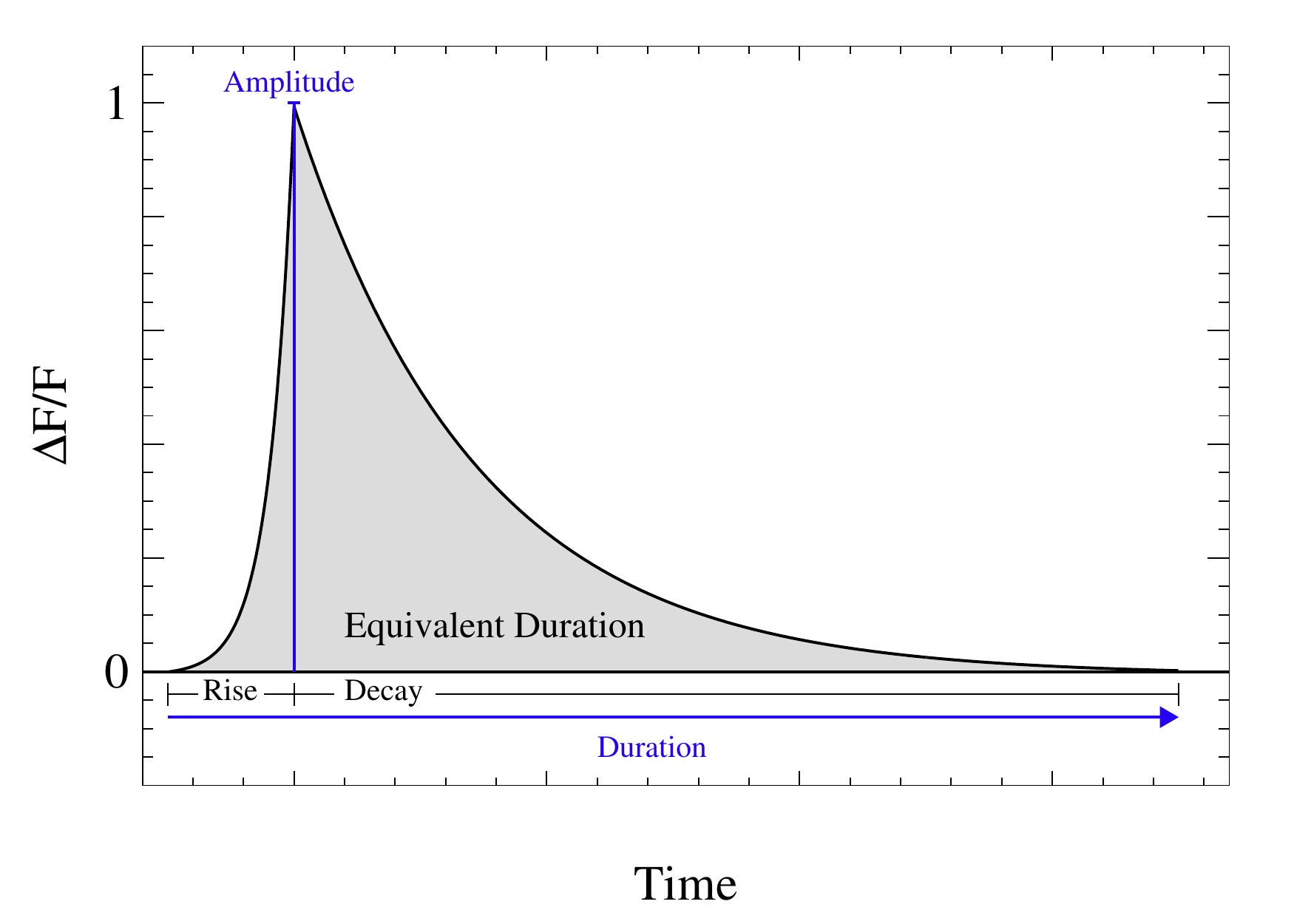}
\caption{Idealized classical flare light curve showing measured quantities -- amplitude, rise and decay times, duration and equivalent duration.}
\label{fig:samplelc}
\end{figure}

The shape of a typical flare light curve (the light curve temporal morphology) is discussed in detail for GJ 1243 in Paper 2.  Here we simply show in Figure \ref{fig:samplelc} an idealized, classical (only one peak) flare with the usual fast rise and fast decay (impulsive phase) followed by a slower exponential decay (gradual phase) light curve, to illustrate the measurements we obtain directly from the Kepler data for each flare -- the amplitude, rise and decay times, duration and equivalent duration.  The amplitude of the flare is defined by the point in time with the highest
flux value (the flare peak), and is expressed as a fraction of the local mean flux, 
\begin{equation}
Amplitude  = (F_{peak}-F_{local mean}) / F_{local mean}
\end{equation}
similar to the fractional flux $\Delta F/F$ defined for the light curves (\S 2.1, Equation 2) but using the local mean flux rather than the global median flux as the reference.  The amplitude of
the flare is thus measured relative to the current state of the underlying star, including effects from starspots, and represents the excess emission above the local mean flux.  For example, the idealized flare shown with an amplitude of 1 in Figure \ref{fig:samplelc} corresponds to a peak flux enhancement that is twice as large as the local mean flux. 
The duration of the flare is the difference between the start time (the point in time when the flare flux begins to deviate from the local mean flux) and the end time (when the flare flux returns to the local mean flux). The start and end times for each vetted flare were obtained during the manual inspection step, including any point where at least two people agreed it belonged in the flare.  The rise time is then the time between flare start and flare peak, the decay time is the time between flare peak and flare end, and the duration is by definition the sum of the rise and decay times. The measured rise and decay times and durations are given in integer minutes due to the one-minute sampling in the Kepler data.  Finally, the equivalent duration is the area under the flare light curve, i.e. the fractional flux integrated over the flare duration as illustrated by the shaded area in Figure \ref{fig:samplelc}.  It is directly related to the flare energy as described in the next section.

 We note that the time sampling and the presence of noise in the light curve means that the detection of the first point in the initial flare rise, and the last point in the flare decay are typically underestimated compared to an analytical flare model.  However, comparison with the model for classical flares discussed in Paper 2 shows that the main effect is an underestimation of flare decay time and duration by about 20\%, and this is independent of the size of the flare.  Further, since the empirical measurements effectively cut off only the last part of the long exponential tail, there is almost no effect on the energy of the flare.  We have chosen to use the empirical measurements in this paper, particularly so that complex and classical flares can be uniformly compared.
 
The quarter of Kepler data used, the total length of the one-minute cadence data
that were analyzed, and the number of flares found for
each target are listed in Table 2.  The quiescent Kepler luminosity and
the range of flare energies found for each object in the sample 
are also given in Table 2 and described in the next section.

\subsection{Determination of Flare Energies}

The flare energy is calculated using the equivalent duration \citep{gershberg1972, huntwalker2012},
which is the area under the flare light curve as described above and shown in Figure \ref{fig:samplelc}.   The equivalent duration is measured in units of time; multiplying it 
by the quiescent luminosity of the star in the Kepler bandpass gives the Kepler flare energy, which we denote as $E_{Kp}$ .

Obtaining the quiescent luminosity in the Kepler bandpass, $L_{Kp}$ 
for each of our
targets required several steps.  First, we
calculate the quiescent Kepler flux (at Earth) using the
measured apparent magnitude $m_{Kp}$ and the
Kepler zero-point magnitude.  To find the zero-point, we convolved
spectrophotometric data for GJ 1243 from \citet{kowalski2013} with the
Kepler filter curve to obtain the specific flux in the Kepler bandpass, 
6.66\e{-14} ergs sec$^{-1}$ cm$^{-2}$ \AA$^{-1}$.  With $m_{Kp}$ = 12.7 for GJ 1243, we solved 
for the Kepler zero-point magnitude 
${Kp_0}$ = -20.24.  The resulting Kepler fluxes for each target were multiplied
by 4$\pi{d^2}$ and the Kepler filter FWHM (4000\AA) to give the 
Kepler quiescent luminosities $L_{Kp}$ shown in Table 2.  As a consistency
check, we convolved the spectra shown in Figure \ref{fig:spec} 
with the Kepler filter curve to obtain measured Kepler fluxes, and found that
these were $\sim$30-80\% of the values obtained from the measured apparent magnitudes using the method outlined above.  This was as expected due to cloud cover and
slit losses during the non-photometric 
conditions when the spectral data were obtained.

A conservative estimate leads to errors of  $\pm$0.2 in the final values
of log $E_{Kp}$, due to uncertainties in the Kepler zero-point, the measured Kepler
magnitudes, the distance estimates and the measurement of the flare equivalent
durations from the Kepler photometry.

\begin{figure}[]
\centering
\includegraphics[width=3.5in]{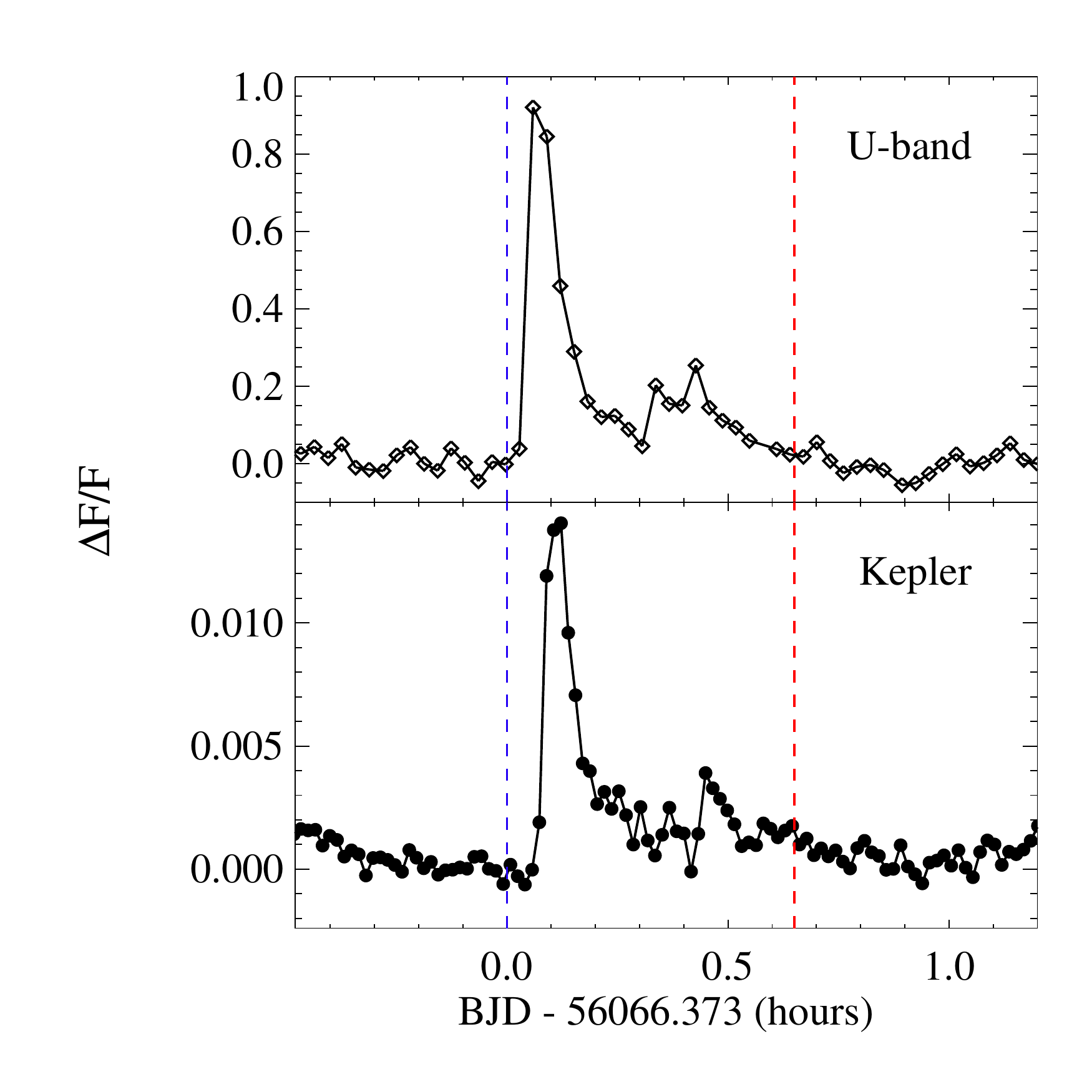}
\caption{Simultaneous Kepler and U-band (from the NMSU 1m telescope at APO) observations of a flare on GJ 1243.
The start and stop times used to compute the equivalent durations are 
indicated by vertical dashed lines.}
\label{fig:kep2u}
\end{figure}

The final step in the data reduction is the transformation 
between the flare energy in the Kepler bandpass and in
a broad-band filter such as the Johnson U-band, traditionally used in flare
studies because flares are quite blue compared to the underlying
stellar light.  However, because equivalent
durations change in different bandpasses, we cannot
simply multiply the Kepler equivalent duration by the quiescent U band
luminosity to obtain $E_U$.  Instead we determined the relationship between
$E_U$ and $E_{Kp}$ using simultaneous data obtained for a flare on GJ 1243
with both Kepler and in the
$U$-band with the NMSU 1-meter telescope at Apache Point Observatory.  The flare light
curves are shown in Figure \ref{fig:kep2u}.  The equivalent durations differ by nearly two
orders of magnitude, with $ED_U$ = 422 seconds, and $ED_{Kp}$ = 4.4 seconds.
GJ 1243 does not have a published $U$ magnitude, so we adopted
U-B = 0.93 (the value for YZ CMi, a famous nearby flare star of similar
spectral type), to obtain log $L_U$ = 28.50 erg s$^{-1}$.  The flare energies are then
log $E_{Kp}$ = 31.31 and log $E_{U}$ = 31.10, or $E_U$ = 0.65$E_{Kp}$.

The energy scaling between the U-band and Kepler filter may vary between flares
for several reasons.  The strong blue emission which characterizes ground-based observations of flares is emitted primarily during the flare
impulsive phase, which varies in strength and duration \citep[see e.g.][]{kowalski2013}.
Flares with a very strong and long-lived impulsive phase will emit relatively more
energy in the U-band. During the flare gradual phase, \citet{kowalski2013} have recently
discussed the existence of red continuum emission in some flares which will contribute to
the emission in the Kepler filter if present.   Flares that have been observed with many filters simultaneously are rare, but a very large flare observed 
on the dM3e star AD Leo simultaneously in the Johnson UBVR filters was reported in
\citet{slhadleo}.  Their Table 6 indicates that the sum
of the B+V+R filter energies covers the Kepler bandpass from $\sim$4000-8000\AA,
and gives $E_U$ = 0.4($E_B + E_V + E_R$) $\sim$ 0.4($E_{Kp}$).
Flare energy scaling relations have also been
determined by LME (and confirmed by Hilton 2011) for a few flares
observed simultaneously with U, B and V filters.  LME found that $E_U = 1.2E_B$
and $E_U = 1.8E_V$.  Again using the sum of the B, V and R filters to approximate the Kepler filter, these
relations give $E_U = 0.6E_{Kp}$ if $E_R = 0.5 E_V$ or $E_U = 0.4E_{Kp}$ if
$E_R = 2 E_V$ (the value found in \citet{slhadleo} for the large AD Leo flare).  Clearly, the presence
and strength of a
red continuum component in a flare will strongly affect the resulting energy scaling.

For our purposes, we have chosen to adopt the
scaling $E_U = 0.65E_{Kp}$ determined from the direct measurement.  However,
variations in the scaling factor from 0.4 to 0.65 lead to a range of $\sim$0.2 in log $E_U$,
and we display this range in our discussion in the next section.

\section{Flare Frequency on Active and Inactive M Dwarfs}

\begin{figure}[]
\centering
\includegraphics[width=3.5in]{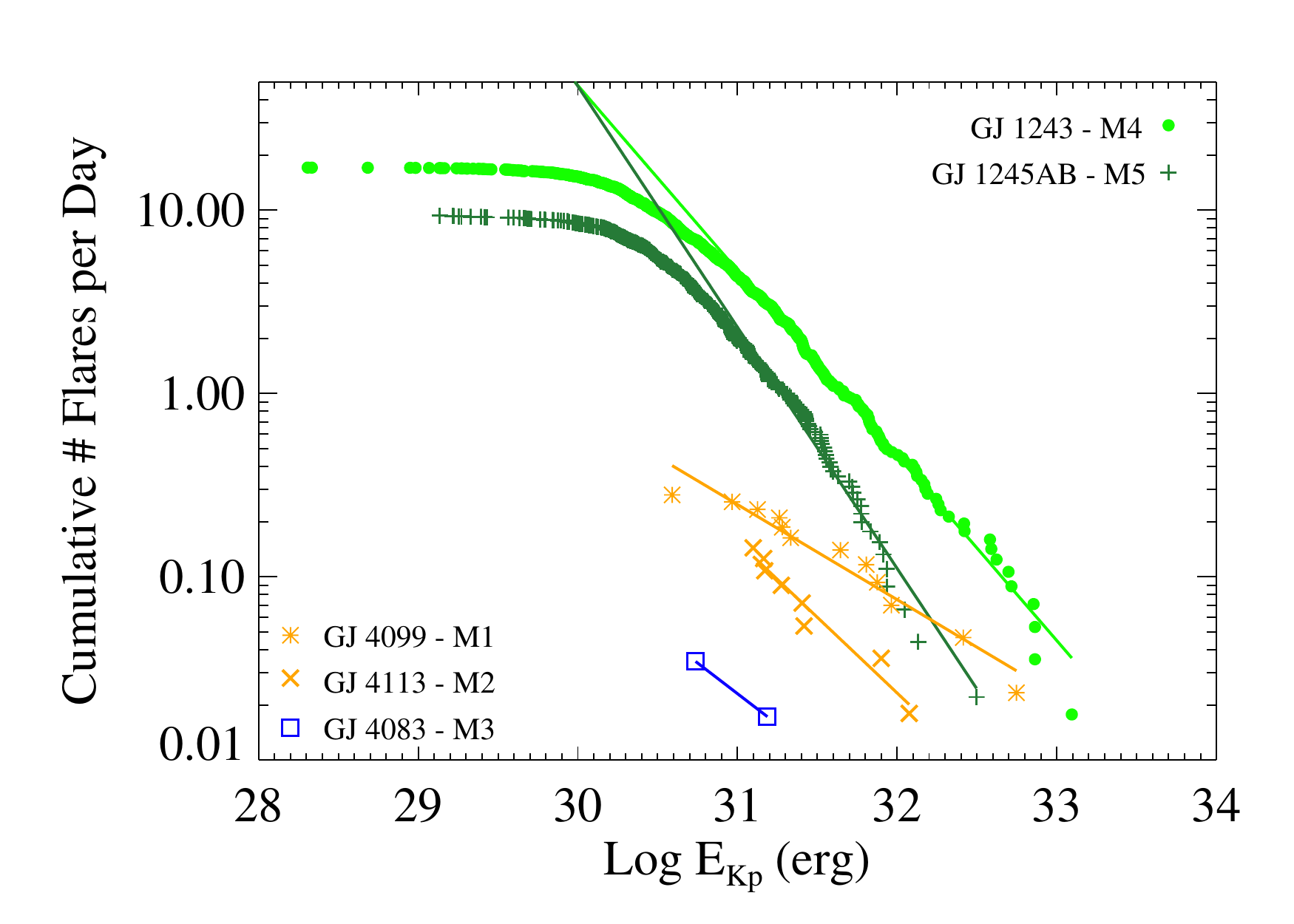}\\
\includegraphics[width=3.5in]{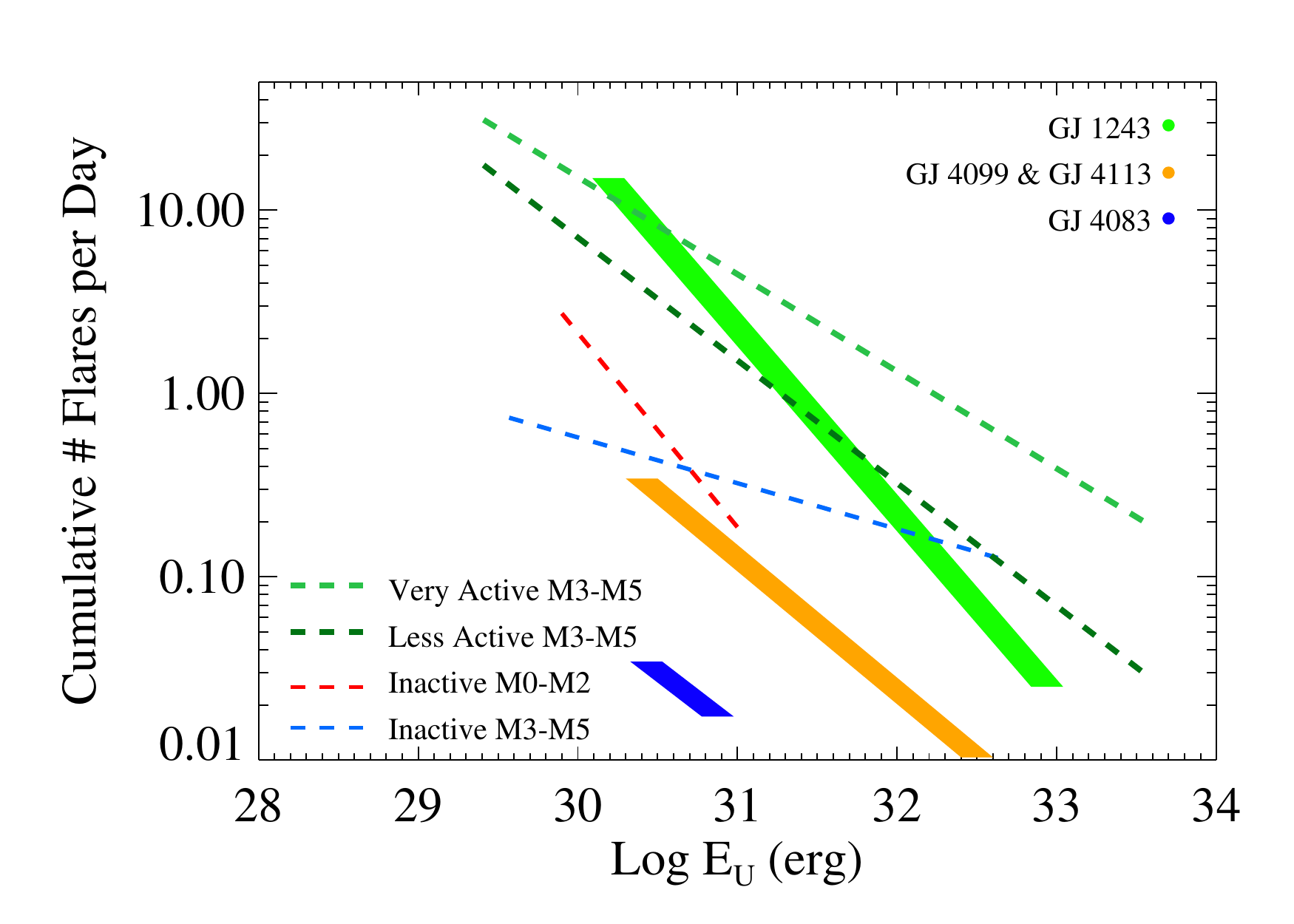}
\caption{The top panel shows the cumulative flare frequency distributions, together with least squares power-law fits, for the active and inactive M stars in the 
Kepler sample, using the native Kepler flare energies.   The bottom panel 
uses the Kepler to U-band energy scaling from \S 2.2 to compare the Kepler
results with ground-based data.  The mean power law relations for nearby M dwarfs from \citet{hiltonthesis} are shown with dashed lines.  See text for discussion.}
\label{fig:ffd}
\end{figure}

Figure \ref{fig:ffd} illustrates our observed flare frequency distributions (FFDs), which are diagrams of cumulative flare frequency (log number of flares per day with energy greater than E) 
as a function of flare energy (log E).\footnote{This diagram is similar to
those shown in LME but with the energy as the independent variable, following more recent studies \citep{audard2000, hiltonthesis, ramsay2013}.}  The 
results in the Kepler bandpass for the five objects in our sample are given in the
top panel, while the bottom panel shows the Kepler data transformed into U-band energies
and compared with previous ground-based studies.  

It is common to characterize the cumulative flare frequency as a power law in energy, for energies
above some limiting energy set by the completeness of the sample.
The low energy
turnover seen most clearly in the active stars may be partially a result of incompleteness in the
detection of small flares and is excluded from the fit. (In \S 4.1 below, we further
discuss this turnover at low energies, using the large sample of flares observed 
on GJ 1243.)  GJ 1243 has a very high flaring rate, comparable to those seen
on the most active, well-observed nearby flare stars such as AD Leo and YZ CMi.
The power-law fit (green solid line) to the GJ 1243 data shown in the top panel of Figure \ref{fig:ffd} was obtained from a least squares fit to the high energy flares (using Poisson uncertainties so that the very highest energy flares do not receive undue weight), increasing the low energy limit until the slope of the power law did not change within the uncertainty in the fit.
This resulted in a limit of log $E_{Kp}$ = 31, meaning that flares of this energy and above are used to fit the power law prediction for flare frequency. 

The GJ 1245AB flares are
at somewhat lower energy, which is expected since the stars are of slightly later spectral type, as discussed in LME.  It is surprising that 
the flare rate is reduced compared to GJ 1243 since flares from both stars in the GJ 1245AB system are
being measured.  In addition,
GJ 1245A is rotating significantly faster than GJ 1243.  A simple
activity-rotation relation would predict stronger activity and more flares on
GJ 1245A.  However, much of the light from GJ 1245A was excluded from the Kepler photometric aperture,  the threshold for detecting flares is higher because the background flux comes from (parts of) both stars, and the signal-to-noise ratio in the GJ 1245AB data is lower, as described in \S 2.1.  All of these factors add uncertainty to the analysis of the GJ 1245AB data from the standard Kepler processing pipeline.  While the combined data for GJ 1245AB qualitatively indicate a somewhat steeper power law than for GJ 1243, we defer quantitative discussion of the individual FFDs to Paper 3, where we investigate this system in more detail, using individual light curves for each component obtained from the Kepler pixel files. 

The inactive stars show significantly lower flare frequency, but it is notable that they
still exhibit some energetic flares.  Evidently, "inactive" as measured by the
lack of H$\alpha$ emission does not mean that these stars have no magnetic field.
Rather, it appears that the field is weaker and/or less organized into substantial active regions
where magnetic heating can produce a strong, persistent chromosphere.  Despite 
the lack of a strong chromosphere, magnetic energy is still being stored and released
periodically in flares which appear similar to the flares seen on the active stars.  The FFDs for the inactive stars suffer from poor sampling (small numbers), especially for the mid-M star GJ 4083; linear least squares fits are shown, which are useful mainly to characterize the general region populated by these stars in the FFD.  
The power-law slopes and resulting values for $\alpha$ (the exponent in the flare energy probability distribution, see Equation 1) for the four single stars in our sample are given in Table 3.

The bottom panel  of Figure \ref{fig:ffd} shows the Kepler data transformed into U-band energies
and the comparison with previous ground-based studies.  
The GJ 1243 FFD is illustrated with the green band; the inactive early type stars have been combined into a single FFD (orange band, slope given in Table 3); and the inactive mid-M star (GJ 4083) is shown with the blue band.  The width of the bands indicates
 the uncertainty due to the
energy scaling as discussed in \S 2.3.
The dashed lines are power-law fits to hundreds of hours of monitoring
data in the U band on a large sample of M dwarfs, as reported in 
\citet[][see Figures 4.14 and 4.19]{hiltonthesis}.  The light green dashed line corresponds
to the very active mid-M stars (including YZ CMi and AD Leo), the dark green dashed line to a set
of less active mid-M stars, the red dashed line to the inactive early-M stars, and the blue dashed
line to the inactive mid-M stars.  The slopes of these power-law fits to the ground-based data and resulting values for $\alpha$ are also given in Table 3 for comparison.

The Kepler FFD for GJ 1243 falls between the very active and less active mid-M relations.  However, it is apparent that the Kepler data exhibit a steeper power law, with fewer flares predicted
at high energy.  Steeper power law fits to the FFD from Kepler data have also been found by \citet{ramsay2013} who compared results for two M dwarfs (including GJ 1243) with numerous individual M dwarfs measured from the ground.   One possible explanation is that   
the long consecutive monitoring periods in the Kepler data provide better sampling of the (rarer) high 
energy flares, and thus allow a more accurate estimate of the power law fit in the high energy regime.

The results for the inactive early M stars in the Kepler sample, GJ 4099 (M1) and GJ 4113 (M2) (orange band) exhibit flare
frequencies similar to those found by the extension of the ground-based relation (red dashed line) to higher energies; the longer duration
of continuous monitoring with Kepler resulted in the detection of more flares at higher
energy than previously were seen on this class of stars.  The small numbers in both the Kepler and ground-based samples ($\sim$ 20 flares
in each) mean that both fits suffer from incompleteness, but it is reassuring that they
occupy a similar region in the diagram and plausibly result from measuring the same underlying
distribution using instruments with different sensitivity.

The inactive M3 star GJ 4083 is quite anomalous, with only 2 measured flares, and
a flare frequency distribution (blue band) more 
than a magnitude lower than predicted by the ground-based relation for inactive mid-M stars (blue dashed line) which itself only includes 3 flares, albeit in a much shorter monitoring period.  The flare rates on
inactive mid-M stars apparently can vary by almost two orders of magnitude, although the small numbers of flares observed mean that this class of stars warrants further investigation.  We note that \citet{paulson2006} and \citet{france2012} have observed serendipitous flares on the inactive dM4 stars Gliese 699 and Gliese 876 respectively, although there are no measurements of the flare frequency distributions for those stars.  GJ 4083 is extremely quiet in its flaring rate although it did show a small (erratic) signature of rotational modulation.  It may be a good candidate for inclusion in planet
search programs that aim to target stars with very low levels of magnetic
activity.

Another measure that we can compute is the fractional energy released in flares
compared to the total energy output of the star.  These ratios are included in Table 2 
as $f_{E}$, and clearly show that a higher fraction of the stellar 
luminosity is released as flare energy in the active stars, in agreement with
\citet{pettersen1988} and \citet{kowalski2009}.  The fraction of time spent in 
a flaring state, $f_{flare}$, is also given in Table 2.  The active stars 
spend $\sim$30\% of their time flaring at a level detectable with Kepler, 
while the flaring time for inactive stars varies between $\sim$0.01-1\%.  
These fractions are higher than those found by 
\citet{kowalski2009} in their investigation of flares observed in 
SDSS imaging data, probably due to the much higher flare threshold (0.7 mag enhancement in the SDSS $u$ filter) adopted in that study.  

In summary, our Kepler results broadly agree with ground-based data, although we find a steeper power-law fit for the active mid-M star,
perhaps due to better sampling of high energy events.  We suggest that the separation of stars into only two groups
(active and inactive) based on their H$\alpha$ emission is inadequate to describe the full range of
flare activity observed. The very active and less active stars span more than a magnitude in flare frequency, and the (relatively) inactive stars cover an additional 1-2 magnitudes at lower frequency.
This continuum in flaring activity across the entire range of early-mid M dwarfs has important implications for modeling flare occurrence in large surveys
such as LSST \citep[e.g.][]{hiltonthesis}, and for predicting the effects of flares on planets orbiting low mass
stars.

\begin{deluxetable*}{lccccccc}
\tablecolumns{8}
\tablecaption{Kepler Flare Energies and Statistics} 
\tablehead{
	\colhead{Star}&
	\colhead{\# Flares}&
         \colhead{Quiescent log $L_{Kp}$} &
	\colhead{Range (log $E_{Kp}$)} &
	\colhead{Quarter} &
	\colhead{duration} &
	\colhead{$f_{flare}^\dagger$} &
	\colhead{$f_{E}^\star$} 
	}
\startdata
GJ 4099 &  12  & 31.64   & 30.6--32.7  &  Q8  &  42.9d & 0.015 & 7.2\e{-6} \\ 
GJ 4113 &  8  & 31.53   & 31.1--32.1  & Q7 & 55.7d & 0.0056 & 1.9\e{-6} \\
GJ 4083 & 2  &  31.18   & 30.7--31.2  & Q7  & 58.0d & 0.0005 & 2.7\e{-7}   \\
GJ 1243 &  833  & 30.67  &  28.3--33.1  &  Q6  & 56.4d & 0.36  & 5.1\e{-3} \\ 
GJ 1245AB &  450 & 30.22  & 29.1--32.5  &  Q8  & 45.3d & 0.27 & 4.2\e{-3} 
\enddata
\tablenotetext{$\dagger$}{$f_{flare}$ is the fraction of time spent in a flaring state.}
\tablenotetext{$\star$}{$f_{E}$ is the fraction of the star's total energy output released in flares.}
\label{flaretable}
\end{deluxetable*}

\begin{deluxetable}{llcc}
\tablecolumns{4}
\tablecaption{Power law fits to FFDs from Kepler and the Ground}
\tablehead{
	\colhead{Star/Sp Types}&
	\colhead{source}&
	\colhead{power law slope}&
	\colhead{$\alpha$}
	}
\startdata
GJ 4099 - M1 & Kepler$^\dagger$ & -0.52  & 1.52 \\
GJ 4113 - M2 & Kepler & -0.83 & 1.83 \\
Inactive M1-M2  & Kepler & -0.72 & 1.72\\
Inactive M0--M2 &  Ground$^\star$ & -1.06 & 2.06 \\
GJ 4083 - M3 & Kepler & -0.67 & 1.67 \\
Inactive M3--M5 & Ground & -0.25 & 1.25 \\
GJ 1243 - M4 & Kepler & -1.01 & 2.01 \\
Very Active M3--M5 & Ground & -0.53 & 1.53 \\
Less Active M3--M5 & Ground & -0.67 & 1.67 \\
GJ 1245AB - M5+M5 & Kepler & -1.32 & 2.32
\enddata
\tablenotetext{$\dagger$}{Kepler results from this paper, see Figure \ref{fig:ffd}.}
\tablenotetext{$\star$}{Ground-based results from \citet{hiltonthesis}, Table 4.3 ("Corrected" values for the slope $\beta$).}
\label{datatable}
\end{deluxetable}

\section{Basic Flare Properties}

The exquisite precision and very long duration of consecutive observations provided by the Kepler data allow us to analyze some basic properties of flares with unprecedented clarity.  We have chosen to concentrate on the sample of nearly 1000 flares from two months of monitoring on the single star GJ 1243 in our analysis.   We examine correlations among the basic observed parameters - amplitude, duration and energy - and discuss the completeness of the observations at low energy.
We then investigate correlations of rise time and decay time with the flare duration.

\subsection{Flare Amplitude, Duration and Energy}

\begin{figure*}[]
\centering
\includegraphics[width=6in]{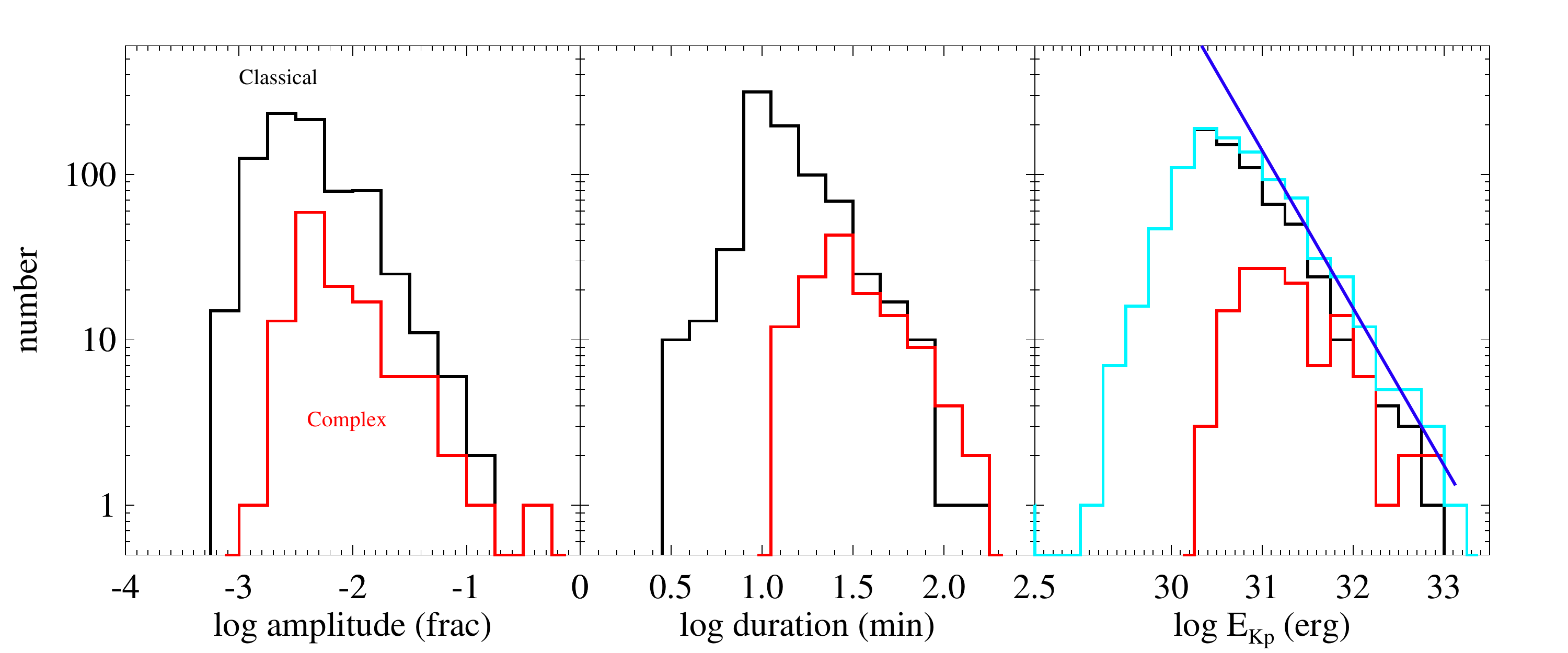}
\caption{The observed distributions of classical (black) and complex (red) flares in amplitude, duration and log energy from two months of monitoring the active M4 star GJ 1243 with Kepler.  The full sample (classical and complex) is shown for the energy diagram (light blue) and fit with a power law for energies above log $E_{Kp} = 31$.}
\label{fig:hist}
\end{figure*}

\begin{figure*}[]
\centering
\includegraphics[width=6in]{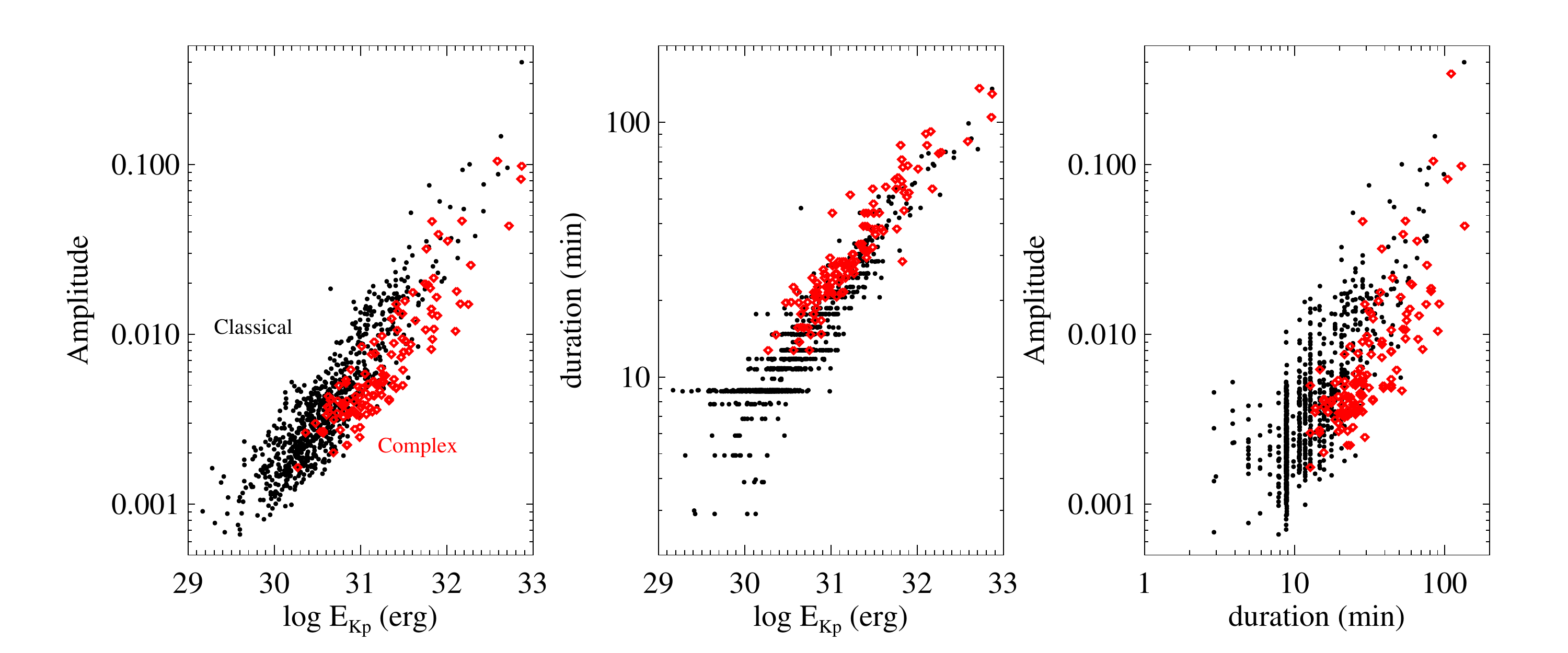}
\caption{Relationships between amplitude, duration and energy for classical and complex flares on GJ 1243.  Larger flare energy is strongly correlated with longer duration, and is also correlated with higher amplitude.  Most high energy flares are complex, but there are still some classical flares even near the highest energies.  The amplitude and duration relationship shows more scatter but still exhibits a significant correlation, with longer flares typically having higher amplitude. }
\label{fig:ampdur}
\end{figure*}

Figure \ref{fig:hist} shows histograms of amplitude, duration and energy for the flares classified as classical (only one peak) and complex (more than one peak).  The strong cutoff in the number of complex flares of short duration is not real, but reflects the one minute time resolution of the Kepler data which makes it difficult to confidently identify multiple peaks in flares shorter than $\sim$ 15 minutes.  The figure shows that classical and complex flares have similar amplitude distributions but complex flares tend to last longer and be more energetic. 

If all flares exhibited an identical light curve, such as the classical flare shown in Figure \ref{fig:samplelc}, then we would expect these measured parameters to be highly correlated, subject only to measurement uncertainties.  In fact, Figure \ref{fig:ampdur} illustrates that both the classical and complex flares do show strong correlations between amplitude, duration and energy.  Note that the one-minute cadence of the observations leads to the artificial integer binning of the duration measurements; we expect that the durations would form a smooth distribution as do the amplitudes and energies if measured with better time resolution.  
In Paper 2, we analyze the light curve morphology using an even larger (11 month) sample of classical flares on GJ 1243 and show that there is a remarkable similarity in classical light curve shape that can be tied to physical emission processes during flares.
Briefly, the rise phase represents a period of rapid impulsive heating characterized by strong blackbody-like emission, while the decay phase first shows an initial rapid decline in the blackbody-like emission followed by a longer slow decline representing gradual cooling from enhanced Balmer and other chromospheric emission.
The regions of blackbody and Balmer emission may also be spatially distinct \citep{kowalski2012}.  \citet{kowalski2013} describes the spectroscopic evidence leading to this interpretation, and the Kepler photometric data presented here and in Paper 2 provide additional evidence that classical flares are well described by such a model.

A power law fit to the energy histogram (above log $E_{Kp} $= 31, found from the FFD, see Figure \ref{fig:ffd}), recovers a slope of -0.95, very close to the slope in the cumulative FFD (-1.01) as expected. The observations deviate from the power law below log $E_{Kp} \sim 31$  which corresponds to duration about twenty minutes and amplitude about 0.005 in fractional flux units.  A typical flare of this energy is indicated in Figure \ref{fig:activelc}.  However, Figure \ref{fig:activelc} also shows that even flares of duration ten minutes (corresponding to amplitude 0.002 and log $E_{Kp} \sim$ 30.5) are easily identified in the Kepler data, and we do not believe we have missed a significant number of such flares, certainly not more than half of them as would be suggested by Figure \ref{fig:hist}.  One possibility is that the lower energy flares do not follow the same power law distribution as the high energy flares (for a discussion of the solar case, see \citet{hannah2011}).  Alternatively, we may not always identify small flares as individual events because they overlap in time with another flare and the entire event is counted in our sample as a single complex flare of higher energy and longer duration.  Disentangling the individual flares in complex events is a difficult problem with data that have no spatial resolution.  In Paper 2, we present results from a Monte Carlo simulation to investigate the incidence of overlapping flares in the 11 month sample for GJ 1243.

The complex flares tend to have lower amplitude but longer duration than classical flares at the same energy, and longer duration and higher energy in general, in keeping with a model where they represent a superposition of classical flares.  
The percentage of flares that are complex increases from 30\% for flares with duration between 20-30 minutes to 60\% for flares with duration more than an hour.  Although we still see some high energy flares that exhibit classical light curves, most high energy (long duration) flares are complex.  They may occur in active regions with complicated magnetic field structures that are capable of storing more energy to be released as flares, and that are therefore more prone to sequential heating of new flaring areas within the same active region, analogous to the appearance of transient optical/UV foot-points during flare evolution as seen in high spatial resolution solar observations \citep{kosovichev2001, wang2007}.  Sympathetic flaring may also occur in nearby active regions, as has also been seen on the Sun \citep{pearce1990}.  However,
longer duration flares are also more likely to randomly overlap with another flare from an independent active region.   Paper 2 addresses the makeup of complex flares using an empirical (classical) flare light curve to decompose the observed complex events, in order to determine the relative importance of these two scenarios, which may both play a role in forming flares with multiple peaks.

\subsection{Flare Rise and Decay Times}

\begin{figure*}[]
\centering
\includegraphics[width=6in]{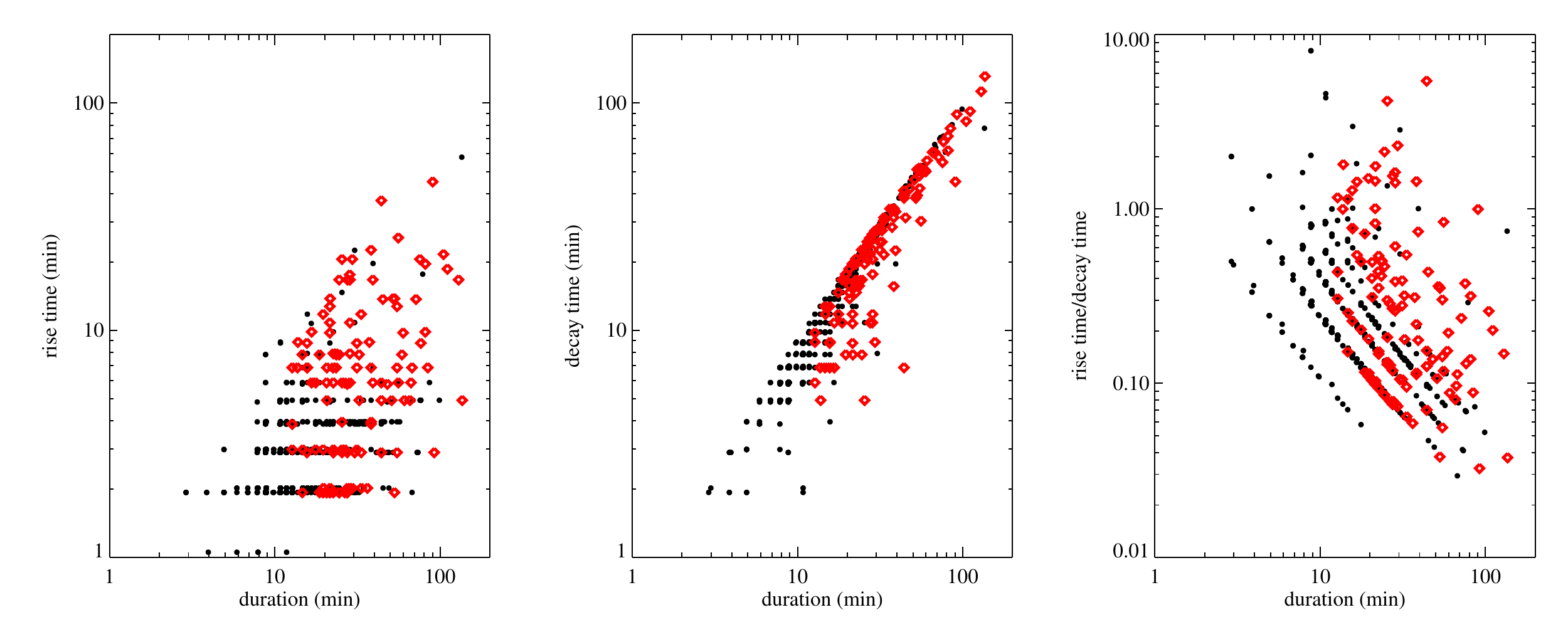}
\caption{The flare rise time is not well correlated with the duration (amplitude, energy) while the decay time is very strongly correlated.  Flares clearly spend the majority of their evolution in the decay phase, which dominates the energy budget.  Complex flares typically have longer rise times at the same duration.  
The ratio of the rise and decay times decreases with duration for classical flares, but complex flares, especially those of long duration, show more complicated behavior (see text). 
 }
\label{fig:risedecay}
\end{figure*}

Figure \ref{fig:risedecay} shows the rise and decay times as a function of flare duration (energy).  The rise time is nearly always shorter than the decay time, and has little correlation with the duration, while the decay time is strongly correlated with the duration.  The flare evolution typically includes a fast rise phase and a longer decay phase, so it is natural that the decay phase dominates the duration.  Since the duration is also strongly correlated with the flare energy, it is clear that most of the energy must be emitted by the flare during the decay phase. Complex flares have longer rise times than classical flares of the same duration, perhaps indicating that multiple events are adding to produce the strongest peak in the light curve, where the rise time is measured. 
The ratio of the rise and decay times shows a correlation with flare duration such that longer duration flares spend relatively more time in the decay phase, particularly for the classical flares.  The ratio is not constant, indicating that the rise phase and decay phase timescales do not simply scale together in a linear fashion.  In fact, the canonical flare morphology that we find in Paper 2 is more complicated, with two distinct decay phases (see discussion in \S4.1).  The longer duration flares are mostly complex, and show little correlation between rise and decay timescales, likely due to multiple heating events contributing to both phases.

\section{Flare Timing}

There are several aspects of flare timing that we can investigate with
the Kepler data on GJ1243, which has both numerous flares and
well-defined starspot modulation.  We first look at the correlation of
flare occurrence (and energy) with starspot phase.  Do flares preferentially
occur when the star is dimmest (largest surface coverage of dark spots on visible
hemisphere)?  Next, we examine the waiting time (time between consecutive flares) distribution.  Do flares tend to cluster together, for example because a new active region has appeared on the surface?  Also, does the flare energy correlate with the energy of the previous flare or the waiting time between flares?  This might have implications for the flare energy buildup and storage mechanisms.

\subsection{Correlation with Starspot Modulation}

\begin{figure}[]
\centering
\includegraphics[width=3.5in]{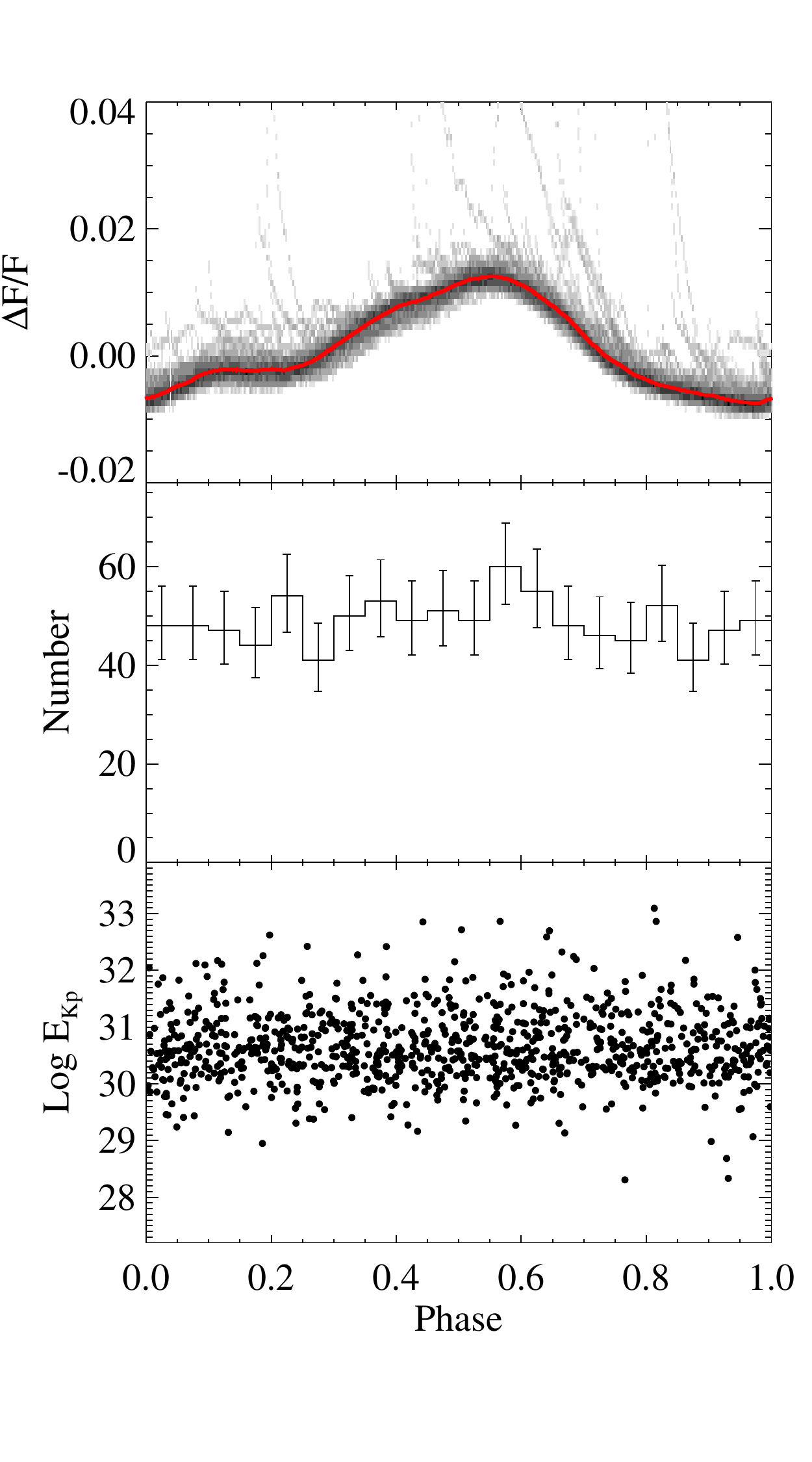}
\caption{The top panel shows one month of GJ 1243 short cadence data folded in phase on the starspot period (0.59 days).  The red line indicates the median value of $\Delta F/F$ at each phase, while the excursions above the median are the flares.  The middle panel illustrates that the number of flares per phase bin is very flat, consistent with no trend with reduced $\chi^2 = 1.1$.   The bottom panel shows similarly that there is no correlation of flare energy with phase.}
\label{fig:flarephase}
\end{figure}

Figure \ref{fig:flarephase} shows the GJ 1243 light curve folded on 
the 0.59 day period determined from the photometric 
modulation which we attribute to starspots on the stellar surface.  
The flares appear as positive flux excursions above the stable, periodic
variation.  The number of flares as a function of phase 
is shown in the 
histogram and is consistent with no trend, with reduced $\chi^2 = 1.1$.  The flare energy is also randomly distributed, and shows no significant
correlation with starspot phase. 

We are carrying out extensive models of the starspot distribution on GJ 1243 (Davenport et al, in prep.) which suggest that an asymmetric polar spot (or more likely spot group) covering a significant fraction of the pole can explain the primary feature of the observed rotational modulation in the light curve.  Since 
a large fraction of the polar spot group will always be in view, we do not expect
a strong correlation of flare occurrence with phase, if the flares are associated with
this spot group.  Alternatively, the flares may be coming from many, relatively small, active regions that contain most of the magnetic flux from the star \citep{reiners2009}, while the polar spot is associated with the weaker dipolar field \citep{morin2008b}.  These many small spots would not contribute
significant modulation compared to that from the large polar spot group, and thus, again, we would not expect a correlation of flare activity with starspot phase.  We cannot distinguish between these scenarios with our existing data.

\subsection {Flare Waiting Time Distribution}

The flare waiting time is often used in solar studies, and defined as the time interval between the start of one flare and the start of the next flare.  We use the entire two-month sample of flares on GJ 1243 in our analysis, including both classical and complex flares.  However, note that complex flares are counted as a single flare event, so the times between individual peaks in complex flares are not considered in our analysis.   

\begin{figure}[]
\centering
\includegraphics[width=4in]{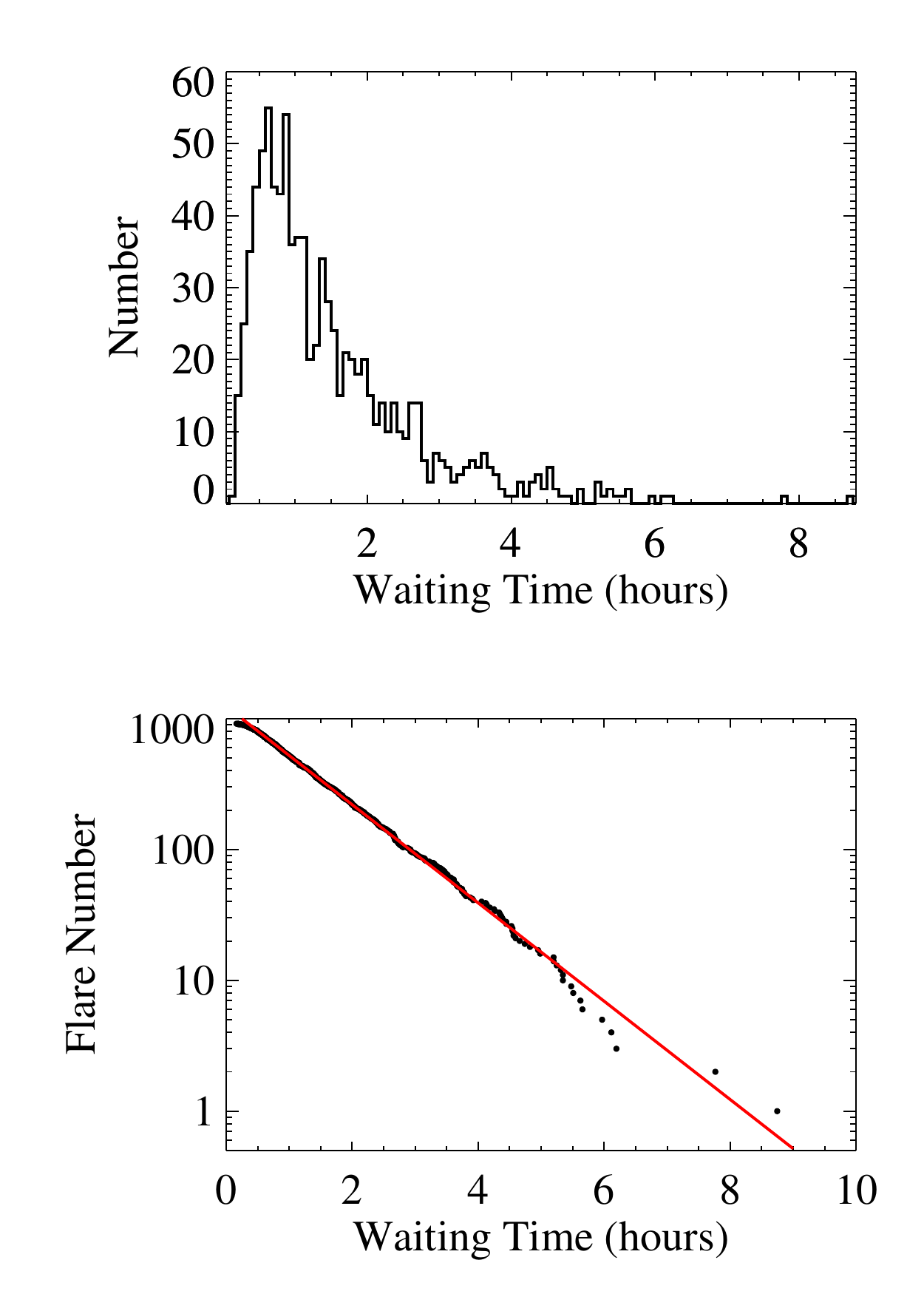}
\caption{Flare waiting times (time between successive flares) measured for GJ 1243 are shown in the histogram (top panel) and as a cumulative distribution (bottom panel).  A single exponential fits the data well above a wait time of 30 minutes.  Flares at smaller wait times are underrepresented, likely because many of them are counted as part of complex events.}
\label{fig:waittime}
\end{figure}

The top panel of Figure \ref{fig:waittime} shows the distribution of waiting times in the GJ 1243 sample.  The decline in the number of flares observed with waiting times below 30 minutes depends on the flare duration distribution, since a new independent flare cannot be identified until the previous flare duration is reached.   Flares that occur while another flare is in progress are considered part of a single complex flare.  The bottom panel gives the cumulative distribution, with the longest waiting time in the sample being almost nine hours; a few flares have waiting times of five hours or more; and all flares have waiting times more than a few minutes (a limit set by the time resolution of the Kepler data and the duration of the shortest flares in the sample).  The cumulative distribution is well fit with a single exponential function (with slope -0.375) for waiting times between 30 minutes and five hours, while the extension of the fit to longer waiting times is adequate given the small numbers.  The data are consistent with a model where flares occur with a single, random, Poisson distribution of waiting time for the entire extent of our observations.   In contrast, \citet{wheatland2010} investigated the evolution of a single solar active region over a weeklong period, and found a piecewise Poisson distribution of waiting times, with an initial flaring period having a shallow slope (longer time between flares) followed by an active flaring period with short waiting times and steeper slope, and then a decay period again with longer waiting times and shallower slope.  

Our interpretation is that GJ 1243 maintains a steady state of flaring over long time periods (months) compared to both its rotation period (fraction of a day) and compared to the behavior of active regions on the Sun (weeks).  This is consistent with a scenario where there are numerous small spot regions (clustered in a large polar spot group, and/or spread out over the surface) that are constantly emerging and decaying.  At any one time, a large number of regions are contributing to the overall flare rate so that while some may be in the buildup or decay phases seen on the Sun, many others are in the strongly flaring phase.  The sum of all this flaring activity thus remains relatively constant with time.  We are investigating the flare frequency over the entire eleven months of short cadence data on GJ 1243 in the Kepler database and indeed it appears quite similar on monthly timescales (Wisniewski et al., in prep.), which supports this steady state scenario of flare activity on GJ 1243. 

\subsection{Flare Energy Distribution}

\begin{figure}[]
\centering
\includegraphics[width=3.4in]{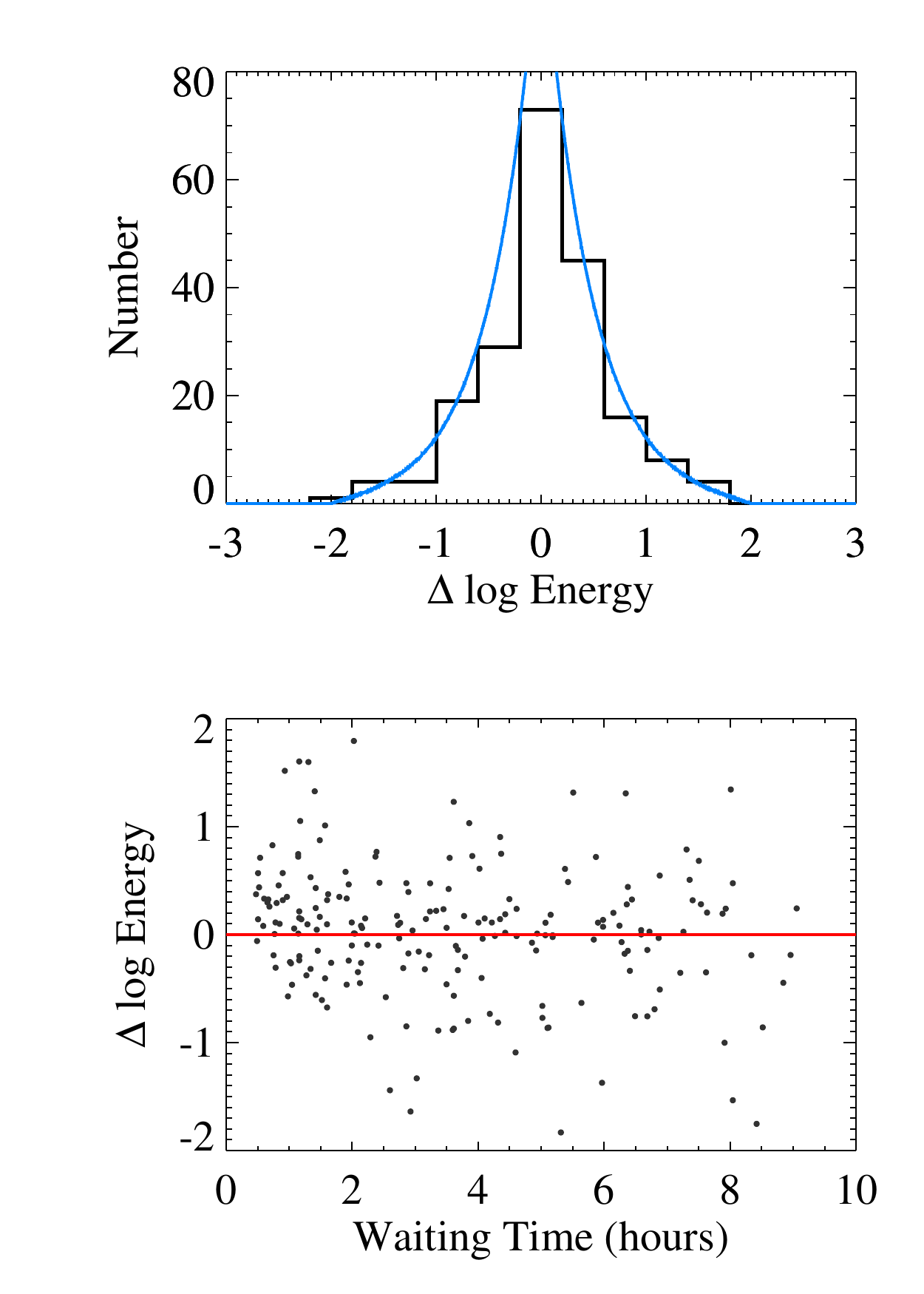}
\caption{The top panel shows the distribution of $\Delta$ log E between successive flares (black histogram).  The model (blue line) gives the expected distribution if flares are drawn randomly from the power law fit to the GJ 1243 flare energies, and is consistent with the observed distribution.  The bottom panel shows that the energy difference between successive flares is not correlated with the waiting time.  Only flares with energy log $E_{Kp}$ $>$ 31 are included.}
\label{fig:energy}
\end{figure}

We define $\Delta$ log E as the energy difference between subsequent flares, 
i.e. $\Delta$ log E = log E$_{i+1}$ - log E$_i$, which is positive if a flare is more energetic than the previous one.  The distribution of $\Delta$ log E for the full sample of GJ 1243 flares, including both classical and complex flares with log E $ >$ 31, is shown as the histogram in the top panel of Figure \ref{fig:energy}.  The flare energies appear to be randomly distributed, with no preferred scale or sign for $\Delta$ log E. 
The smooth blue line represents a model distribution of energy differences obtained by choosing flares randomly from a power-law flare energy distribution with $\alpha = 1.95$, found from the
fit to the energy histogram of all events in \S 4.1. The random model is consistent with the observations of flares with log E $> $ 31, where the power law is applicable.

The distribution of $\Delta$ log E with waiting time is shown in the bottom panel of Figure \ref{fig:energy}.  Again there is no correlation of waiting time with the energy of the subsequent flare.  We do expect, and observe, a region in the lower left of the diagram at negative $\Delta$ log E (first flare is higher energy) and small waiting time that is relatively unpopulated.  This is because larger flares typically have longer duration, and hence necessarily a longer waiting time before the next flare can be unambiguously identified.  

These results support a scenario where the individual flares we have measured are randomly occurring at times and energies that do not depend on the previous flare history.    Thus, each event may be occurring in a separate active region, not triggered by previous events.  We do not find any evidence of sympathetic flaring between individual events, as suggested for example by \citet{panagi1995}.  We stress again that we treat complex flares as single events in this analysis.  Within a complex flare, the sub-peaks that we see may sometimes be triggered from the initial reconnection and energy release, see e.g. \citet{anfinogentov2013}.  We discuss the makeup of complex flares in more detail in Paper 2.

\section{Summary}

We have analyzed two months of Kepler short-cadence data on several low mass M dwarfs, spanning a range of activity and spectral type.  We obtained additional spectroscopic data from the ground to assign spectral types, and used simultaneous Kepler and U-band photometry of one flare to obtain an energy-scaling relationship between Kepler and U-band flare energies.  We report periods for five stars; these agree with published periods for the two stars that have previous determinations.

The flare frequency distributions (FFDs) from Kepler data probe higher flare energies with better statistics than are available from the ground.  We suggest that this leads to steeper power-law fits to the FFDs compared to those found from ground-based data, indicating that the power law energy distribution of flares may approach $\alpha =2$, with implications for coronal heating.  However, we also find evidence that the power law fit does not continue to lower energies, which are still above our detection limit.  The missing low energy flares may be subsumed into complex events, or there may be a real turnover in the power law energy distribution of flares. We investigate these possibilities further in Paper 2 of this series.

Comparison of the FFDs for inactive and active stars of early and mid-M spectral type shows that there is a continuum of flare frequencies and energies, and that the artificial separation into "active" and "inactive" classes based solely on the presence or absence of H$\alpha$ emission in the quiescent spectrum does not capture the full complexity of magnetic (flaring) activity in M dwarfs.  Flares from inactive stars, though relatively rare, can be of high energy, and the large number of early-type, inactive M dwarfs in the Galaxy means that these flares could be an important source of transient phenomena in time domain surveys.  In addition, flares from inactive stars may still play an important role in the atmospheric chemistry of orbiting planets.

We also examined the properties of basic flare parameters for a sample of nearly 1000 flares on the active M4 star GJ 1243, and found that flare amplitude, duration and energy are all strongly correlated.  High energy flares typically have both long duration and high amplitude.  Complex flares (with more than one peak) have longer duration and higher energy at the same amplitude.  The fraction of flares that are complex also rises with flare duration.  These results support a scenario where complex flares represent the superposition of classical flares, possibly from flares in separate regions that randomly overlap in time and/or from triggered flaring within the same or nearby active regions.  Paper 2 continues the exploration of the makeup of complex flares using a classical flare model to decompose the complex flare light curves.

The precision and long monitoring period of the Kepler data also allowed us to investigate the correlation of flare occurrence and energy with starspot phase on GJ 1243.  We found no significant correlation, indicating that the spot or spot group that contributes to the periodic photometric modulation does not represent the only, or even the primary, spatial location where flares occur. Instead, the random occurrence of flares with phase indicates that flaring is distributed across the stellar surface as it rotates, possibly in many small active regions.  These results are consistent with the large scale magnetic structure model of \citet{morin2008b}, which accounts for only a small fraction of the total magnetic energy in active mid-M stars; most of the energy is likely found in a large number of smaller active regions that are not detectable in global polarization measurements but contribute to the large magnetic field measurements of several kG with large filling factor \citep{cmj1996}.

Finally, we considered the flare waiting time distribution, and found that it is well described with a single exponential, indicating a random Poisson distribution of flare waiting times above 30 minutes.
The evolution of a single active region on the Sun shows a piecewise Poisson distribution, with periods of weak flaring as the region builds and decays, and strong flaring in between \citep{wheatland2010}.  We interpret the steady-state flaring distribution that persists for the entire two months on GJ 1243 as additional evidence for the contribution of many active regions in a strongly flaring state, such that a few regions that are in the buildup and decay phases do not contribute significantly to the overall rate.  We do not find evidence for sympathetic flaring, but note that complex events are treated as single flares in our analysis.  Sympathetic flaring within a given complex flare may still be occurring.

The difference in energy between successive flares was also investigated, and the results were consistent with a model where flares are randomly chosen from a power-law energy distribution with $\alpha$ appropriate for GJ 1243.  This does not support the notion of independent precursor flares, where a large flare is immediately preceded by a smaller flare.  The waiting time between flares also showed no correlation with the energy of the next flare, indicating that the flares are occurring in independent regions, or that the magnetic energy stored in a given region is much larger than is being emitted in flares.

In summary, our data support a model where many small-scale active regions are distributed across the stellar surface, contributing only marginally to the rotational modulation of the star, and flaring independently at random intervals, with no connection in time or energy between successive flares.

Additional studies using our Kepler short cadence data will investigate the temporal morphology of the classical flare light curve and the makeup of complex flares (Paper 2); the flaring behavior of the individual components of the GJ 1245AB system (Paper 3); the flare frequency and multi-wavelength behavior of flares in the full 11-month Kepler dataset, together with ground-based data, on GJ 1243 (Wisniewski et al., in prep.); and the comparison of flares on stars of types G, K and M (Hawley et al., in prep.).  The Kepler satellite has delivered stellar light curves of unprecedented precision and duration and will provide a rich data source for flare studies for the foreseeable future.

\acknowledgments

This work was supported by Kepler Cycle 2 GO grant NNX11AB71G and Cycle 3 GO grant NNX12AC79G . JRAD acknowledges support from NASA ADP grant NNX09AC77G. SLH, JRAD and LH acknowledge support from NSF grant  AST13-11678.  EJH, AFK, and SLH acknowledge support  from NSF grant AST08-07205.

Observations reported here were obtained with the Apache Point Observatory 3.5-meter telescope, which is owned and operated by the Astrophysical Research Consortium.

This paper includes data collected by the Kepler mission. Funding for the Kepler mission is provided by the NASA Science Mission directorate. Some of the data presented in this paper were obtained from the Mikulski Archive for Space Telescopes (MAST). STScI is operated by the Association of Universities for Research in Astronomy, Inc., under NASA contract NAS5-26555. Support for MAST for non-HST data is provided by the NASA Office of Space Science via grant NNX13AC07G and by other grants and contracts.



\begin{thebibliography}{51}
\expandafter\ifx\csname natexlab\endcsname\relax\def\natexlab#1{#1}\fi

\bibitem[{{Anfinogentov} {et~al.}(2013){Anfinogentov}, {Nakariakov},
  {Mathioudakis}, {Van Doorsselaere}, \& {Kowalski}}]{anfinogentov2013}
{Anfinogentov}, S., {Nakariakov}, V.~M., {Mathioudakis}, M., {Van
  Doorsselaere}, T., \& {Kowalski}, A.~F. 2013, \apj, 773, 156

\bibitem[{{Audard} {et~al.}(2000){Audard}, {G{\"u}del}, {Drake}, \&
  {Kashyap}}]{audard2000}
{Audard}, M., {G{\"u}del}, M., {Drake}, J.~J., \& {Kashyap}, V.~L. 2000, \apj,
  541, 396

\bibitem[{{Bochanski} {et~al.}(2010){Bochanski}, {Hawley}, {Covey}, {West},
  {Reid}, {Golimowski}, \& {Ivezi{\'c}}}]{bochanski2010}
{Bochanski}, J.~J., {Hawley}, S.~L., {Covey}, K.~R., {West}, A.~A., {Reid},
  I.~N., {Golimowski}, D.~A., \& {Ivezi{\'c}}, {\v Z}. 2010, \aj, 139, 2679

\bibitem[{{Bochanski} {et~al.}(2012){Bochanski}, {Hawley}, {Covey}, {West},
  {Reid}, {Golimowski}, \& {Ivezi{\'c}}}]{bochanski2012}
---. 2012, \aj, 143, 152

\bibitem[{{Borucki} {et~al.}(2010){Borucki}, {Koch}, {Basri}, {Batalha},
  {Brown}, {Caldwell}, {Caldwell}, {Christensen-Dalsgaard}, {Cochran},
  {DeVore}, {Dunham}, {Dupree}, {Gautier}, {Geary}, {Gilliland}, {Gould},
  {Howell}, {Jenkins}, {Kondo}, {Latham}, {Marcy}, {Meibom}, {Kjeldsen},
  {Lissauer}, {Monet}, {Morrison}, {Sasselov}, {Tarter}, {Boss}, {Brownlee},
  {Owen}, {Buzasi}, {Charbonneau}, {Doyle}, {Fortney}, {Ford}, {Holman},
  {Seager}, {Steffen}, {Welsh}, {Rowe}, {Anderson}, {Buchhave}, {Ciardi},
  {Walkowicz}, {Sherry}, {Horch}, {Isaacson}, {Everett}, {Fischer}, {Torres},
  {Johnson}, {Endl}, {MacQueen}, {Bryson}, {Dotson}, {Haas}, {Kolodziejczak},
  {Van Cleve}, {Chandrasekaran}, {Twicken}, {Quintana}, {Clarke}, {Allen},
  {Li}, {Wu}, {Tenenbaum}, {Verner}, {Bruhweiler}, {Barnes}, \&
  {Prsa}}]{borucki2010}
{Borucki}, W.~J., {et~al.} 2010, Science, 327, 977

\bibitem[{{Covey} {et~al.}(2007){Covey}, {Ivezi{\'c}}, {Schlegel},
  {Finkbeiner}, {Padmanabhan}, {Lupton}, {Ag{\"u}eros}, {Bochanski}, {Hawley},
  {West}, {Seth}, {Kimball}, {Gogarten}, {Claire}, {Haggard}, {Kaib},
  {Schneider}, \& {Sesar}}]{covey2007}
{Covey}, K.~R., {et~al.} 2007, \aj, 134, 2398

\bibitem[{{Davenport} {et~al.}(2012){Davenport}, {Becker}, {Kowalski},
  {Hawley}, {Schmidt}, {Hilton}, {Sesar}, \& {Cutri}}]{davenport2012}
{Davenport}, J.~R.~A., {Becker}, A.~C., {Kowalski}, A.~F., {Hawley}, S.~L.,
  {Schmidt}, S.~J., {Hilton}, E.~J., {Sesar}, B., \& {Cutri}, R. 2012, \apj,
  748, 58

\bibitem[{{Davenport} {et~al.}(2014){Davenport}, {Hawley}, {Hebb},
  {Wisniewski}, {Kowalski}, \& et~al.}]{davenport2014b}
{Davenport}, J.~R.~A., {Hawley}, S.~L., {Hebb}, L., {Wisniewski}, J.~P.,
  {Kowalski}, A.~F., \& et~al. 2014, \apj, submitted

\bibitem[{{Davenport} {et~al.}(2006){Davenport}, {West}, {Matthiesen},
  {Schmieding}, \& {Kobelski}}]{davenport2006}
{Davenport}, J.~R.~A., {West}, A.~A., {Matthiesen}, C.~K., {Schmieding}, M., \&
  {Kobelski}, A. 2006, \pasp, 118, 1679

\bibitem[{{Donati} {et~al.}(2008){Donati}, {Morin}, {Petit}, {Delfosse},
  {Forveille}, {Auri{\`e}re}, {Cabanac}, {Dintrans}, {Fares}, {Gastine},
  {Jardine}, {Ligni{\`e}res}, {Paletou}, {Ramirez Velez}, \&
  {Th{\'e}ado}}]{donati2008}
{Donati}, J.-F., {et~al.} 2008, \mnras, 390, 545

\bibitem[{{Doyle} {et~al.}(1990){Doyle}, {Butler}, {van den Oord}, \&
  {Kiang}}]{doyle1990}
{Doyle}, J.~G., {Butler}, C.~J., {van den Oord}, G.~H.~J., \& {Kiang}, T. 1990,
  \aap, 232, 83

\bibitem[{{France} {et~al.}(2012){France}, {Linsky}, {Tian}, {Froning}, \&
  {Roberge}}]{france2012}
{France}, K., {Linsky}, J.~L., {Tian}, F., {Froning}, C.~S., \& {Roberge}, A.
  2012, \apjl, 750, L32

\bibitem[{{Friedman}(1984)}]{supersmoother}
{Friedman}, J.~H. 1984, A Variable Span Smoother, Tech. Rep.~5, Department of
  Statistics, Stanford University

\bibitem[{{Gershberg}(1972)}]{gershberg1972}
{Gershberg}, R.~E. 1972, \apss, 19, 75

\bibitem[{{Gizis} {et~al.}(2013){Gizis}, {Burgasser}, {Berger}, {Williams},
  {Vrba}, {Cruz}, \& {Metchev}}]{gizis2013}
{Gizis}, J.~E., {Burgasser}, A.~J., {Berger}, E., {Williams}, P.~K.~G., {Vrba},
  F.~J., {Cruz}, K.~L., \& {Metchev}, S. 2013, \apj, 779, 172

\bibitem[{{G{\"u}del}(2004)}]{gudel2004}
{G{\"u}del}, M. 2004, \aapr, 12, 71

\bibitem[{{Hannah} {et~al.}(2011){Hannah}, {Hudson}, {Battaglia}, {Christe},
  {Ka{\v s}parov{\'a}}, {Krucker}, {Kundu}, \& {Veronig}}]{hannah2011}
{Hannah}, I.~G., {Hudson}, H.~S., {Battaglia}, M., {Christe}, S., {Ka{\v
  s}parov{\'a}}, J., {Krucker}, S., {Kundu}, M.~R., \& {Veronig}, A. 2011,
  \ssr, 159, 263

\bibitem[{{Hawley} \& {Pettersen}(1991)}]{slhadleo}
{Hawley}, S.~L., \& {Pettersen}, B.~R. 1991, \apj, 378, 725

\bibitem[{{Hilton}(2011)}]{hiltonthesis}
{Hilton}, E.~J. 2011, PhD thesis, University of Washington

\bibitem[{{Hudson}(1991)}]{hudson1991}
{Hudson}, H.~S. 1991, \solphys, 133, 357

\bibitem[{{Hunt-Walker} {et~al.}(2012){Hunt-Walker}, {Hilton}, {Kowalski},
  {Hawley}, \& {Matthews}}]{huntwalker2012}
{Hunt-Walker}, N.~M., {Hilton}, E.~J., {Kowalski}, A.~F., {Hawley}, S.~L., \&
  {Matthews}, J.~M. 2012, \pasp, 124, 545

\bibitem[{{Irwin} {et~al.}(2011){Irwin}, {Berta}, {Burke}, {Charbonneau},
  {Nutzman}, {West}, \& {Falco}}]{irwin2011}
{Irwin}, J., {Berta}, Z.~K., {Burke}, C.~J., {Charbonneau}, D., {Nutzman}, P.,
  {West}, A.~A., \& {Falco}, E.~E. 2011, \apj, 727, 56

\bibitem[{{Johns-Krull} \& {Valenti}(1996)}]{cmj1996}
{Johns-Krull}, C.~M., \& {Valenti}, J.~A. 1996, \apjl, 459, L95

\bibitem[{{Kosovichev} \& {Zharkova}(2001)}]{kosovichev2001}
{Kosovichev}, A.~G., \& {Zharkova}, V.~V. 2001, \apjl, 550, L105

\bibitem[{{Kowalski} {et~al.}(2009){Kowalski}, {Hawley}, {Hilton}, {Becker},
  {West}, {Bochanski}, \& {Sesar}}]{kowalski2009}
{Kowalski}, A.~F., {Hawley}, S.~L., {Hilton}, E.~J., {Becker}, A.~C., {West},
  A.~A., {Bochanski}, J.~J., \& {Sesar}, B. 2009, \aj, 138, 633

\bibitem[{{Kowalski} {et~al.}(2012){Kowalski}, {Hawley}, {Holtzman},
  {Wisniewski}, \& {Hilton}}]{kowalski2012}
{Kowalski}, A.~F., {Hawley}, S.~L., {Holtzman}, J.~A., {Wisniewski}, J.~P., \&
  {Hilton}, E.~J. 2012, \solphys, 277, 21

\bibitem[{{Kowalski} {et~al.}(2013){Kowalski}, {Hawley}, {Wisniewski}, {Osten},
  {Hilton}, {Holtzman}, {Schmidt}, \& {Davenport}}]{kowalski2013}
{Kowalski}, A.~F., {Hawley}, S.~L., {Wisniewski}, J.~P., {Osten}, R.~A.,
  {Hilton}, E.~J., {Holtzman}, J.~A., {Schmidt}, S.~J., \& {Davenport},
  J.~R.~A. 2013, \apjs, 207, 15

\bibitem[{{Lacy} {et~al.}(1976){Lacy}, {Moffett}, \& {Evans}}]{lme1976}
{Lacy}, C.~H., {Moffett}, T.~J., \& {Evans}, D.~S. 1976, \apjs, 30, 85

\bibitem[{{McQuillan} {et~al.}(2013){McQuillan}, {Aigrain}, \&
  {Mazeh}}]{mcquillan2013}
{McQuillan}, A., {Aigrain}, S., \& {Mazeh}, T. 2013, \mnras, 432, 1203

\bibitem[{{Moffett}(1974)}]{moffett1974}
{Moffett}, T.~J. 1974, \apjs, 29, 1

\bibitem[{{Morin} {et~al.}(2008){Morin}, {Donati}, {Petit}, {Delfosse},
  {Forveille}, {Albert}, {Auri{\`e}re}, {Cabanac}, {Dintrans}, {Fares},
  {Gastine}, {Jardine}, {Ligni{\`e}res}, {Paletou}, {Ramirez Velez}, \&
  {Th{\'e}ado}}]{morin2008b}
{Morin}, J., {et~al.} 2008, \mnras, 390, 567

\bibitem[{{Morin} {et~al.}(2010){Morin}, {Donati}, {Petit}, {Delfosse},
  {Forveille}, \& {Jardine}}]{morin2010}
{Morin}, J., {Donati}, J.-F., {Petit}, P., {Delfosse}, X., {Forveille}, T., \&
  {Jardine}, M.~M. 2010, \mnras, 407, 2269

\bibitem[{{Osten} \& {Brown}(1999)}]{osten1999}
{Osten}, R.~A., \& {Brown}, A. 1999, \apj, 515, 746

\bibitem[{{Panagi} \& {Andrews}(1995)}]{panagi1995}
{Panagi}, P.~M., \& {Andrews}, A.~D. 1995, \mnras, 277, 423

\bibitem[{{Parker}(1988)}]{parker1988}
{Parker}, E.~N. 1988, \apj, 330, 474

\bibitem[{{Paulson} {et~al.}(2006){Paulson}, {Allred}, {Anderson}, {Hawley},
  {Cochran}, \& {Yelda}}]{paulson2006}
{Paulson}, D.~B., {Allred}, J.~C., {Anderson}, R.~B., {Hawley}, S.~L.,
  {Cochran}, W.~D., \& {Yelda}, S. 2006, \pasp, 118, 227

\bibitem[{{Pazzani} \& {Rodono}(1981)}]{pazzani1981}
{Pazzani}, V., \& {Rodono}, M. 1981, \apss, 77, 347

\bibitem[{{Pearce} \& {Harrison}(1990)}]{pearce1990}
{Pearce}, G., \& {Harrison}, R.~A. 1990, \aap, 228, 513

\bibitem[{{Pettersen}(1988)}]{pettersen1988}
{Pettersen}, B.~R. 1988, in Astrophysics and Space Science Library, Vol. 143,
  Activity in Cool Star Envelopes, ed. {O.~Havnes, J.~E.~Solheim,
  B.~R.~Pettersen, \& J.~H.~M.~M.~Schmitt }, 49--60

\bibitem[{{Ramsay} {et~al.}(2013){Ramsay}, {Doyle}, {Hakala}, {Garcia-Alvarez},
  {Brooks}, {Barclay}, \& {Still}}]{ramsay2013}
{Ramsay}, G., {Doyle}, J.~G., {Hakala}, P., {Garcia-Alvarez}, D., {Brooks}, A.,
  {Barclay}, T., \& {Still}, M. 2013, \mnras, 434, 2451

\bibitem[{{Reid} {et~al.}(1995){Reid}, {Hawley}, \& {Gizis}}]{pmsu1}
{Reid}, I.~N., {Hawley}, S.~L., \& {Gizis}, J.~E. 1995, \aj, 110, 1838

\bibitem[{{Reiners} \& {Basri}(2009)}]{reiners2009}
{Reiners}, A., \& {Basri}, G. 2009, \aap, 496, 787

\bibitem[{{Robinson} {et~al.}(1995){Robinson}, {Carpenter}, {Percival}, \&
  {Bookbinder}}]{robinson1995}
{Robinson}, R.~D., {Carpenter}, K.~G., {Percival}, J.~W., \& {Bookbinder},
  J.~A. 1995, \apj, 451, 795

\bibitem[{{Robinson} {et~al.}(2001){Robinson}, {Linsky}, {Woodgate}, \&
  {Timothy}}]{robinson2001}
{Robinson}, R.~D., {Linsky}, J.~L., {Woodgate}, B.~E., \& {Timothy}, J.~G.
  2001, \apj, 554, 368

\bibitem[{{Schrijver} {et~al.}(2012){Schrijver}, {Beer}, {Baltensperger},
  {Cliver}, {G{\"u}del}, {Hudson}, {McCracken}, {Osten}, {Peter}, {Soderblom},
  {Usoskin}, \& {Wolff}}]{schrijver2012}
{Schrijver}, C.~J., {et~al.} 2012, Journal of Geophysical Research (Space
  Physics), 117, 8103

\bibitem[{{Schroeder} {et~al.}(2000){Schroeder}, {Golimowski}, {Brukardt},
  {Burrows}, {Caldwell}, {Fastie}, {Ford}, {Hesman}, {Kletskin}, {Krist},
  {Royle}, \& {Zubrowski}}]{schroeder2000}
{Schroeder}, D.~J., {et~al.} 2000, \aj, 119, 906

\bibitem[{{Smith} {et~al.}(2012){Smith}, {Stumpe}, {Van Cleve}, {Jenkins},
  {Barclay}, {Fanelli}, {Girouard}, {Kolodziejczak}, {McCauliff}, {Morris}, \&
  {Twicken}}]{smith2012}
{Smith}, J.~C., {et~al.} 2012, \pasp, 124, 1000

\bibitem[{{Stassun} {et~al.}(2011){Stassun}, {Hebb}, {Covey}, {West}, {Irwin},
  {Jackson}, {Jardine}, {Morin}, {Mullan}, \& {Reid}}]{stassun2011}
{Stassun}, K.~G., {et~al.} 2011, in Astronomical Society of the Pacific
  Conference Series, Vol. 448, 16th Cambridge Workshop on Cool Stars, Stellar
  Systems, and the Sun, ed. C.~{Johns-Krull}, M.~K. {Browning}, \& A.~A.
  {West}, 505

\bibitem[{{Walkowicz} {et~al.}(2011){Walkowicz}, {Basri}, {Batalha},
  {Gilliland}, {Jenkins}, {Borucki}, {Koch}, {Caldwell}, {Dupree}, {Latham},
  {Meibom}, {Howell}, {Brown}, \& {Bryson}}]{walkowicz2011}
{Walkowicz}, L.~M., {et~al.} 2011, \aj, 141, 50

\bibitem[{{Wang} {et~al.}(2007){Wang}, {Fang}, \& {Ming-DeDing}}]{wang2007}
{Wang}, L., {Fang}, C., \& {Ming-DeDing}. 2007, Chinese J. Astron. Astrophys., 7, 721

\bibitem[{{Wheatland}(2010)}]{wheatland2010}
{Wheatland}, M.~S. 2010, \apj, 710, 1324

\end{thebibliography}
\end{document}